\documentclass[usenatbib,usegraphicx]{mn2e}

\title[The Low-Redshift Ly$\alpha$ Forest toward 3C~273]{The Low-Redshift Ly$\alpha$ Forest toward 3C~273\thanks{Based
on observations with (1) the NASA/ESA Hubble Space Telescope,obtained at the Space Telescope Science Institute, which is operated
by the Association of Universities for Research in Astronomy, Inc.,
under NASA contract NAS 05-26555, and (2) the NASA-CNES/ESA Far
Ultraviolet Spectroscopic Explorer mission, operated by the Johns
Hopkins University, supported by NASA contract NAS 05-32985.}}

\author[G. M. Williger et al.]{Gerard M. Williger$,^{1,2,3,4}$ Robert F. Carswell,$^5$ Ray J. Weymann,$^6$
\newauthor
Edward B. Jenkins,$^7$ Kenneth R. Sembach,$^8$ Todd M. Tripp,$^9$ Romeel Dav\'e,$^{10}$
\newauthor
Lutz Haberzettl,$^3$ Sara R. Heap$^1$
\newauthor
$^1${\footnotesize Code 667, NASA Goddard Space Flight Center, Greenbelt MD 20771, USA}
\newauthor
$^2${\footnotesize Dept. of Physics \& Astronomy, Johns Hopkins U., Baltimore MD 21218, USA}
\newauthor
$^3${\footnotesize Dept. of Physics \& Astronomy, U. Louisville, Louisville KY 40292, USA}
\newauthor
$^4${\footnotesize present address: Lab. Fizeau, UMR 6525, Univ. de Nice, 06108 Nice Cedex 2, France}
\newauthor
$^5${\footnotesize Institute of Astronomy, Madingley Road, Cambridge CB3 0HA, England}
\newauthor
$^6${\footnotesize Carnegie Observatories, 813 Santa Barbara St., Pasadena CA 91101, USA}
\newauthor
$^7${\footnotesize Princeton U. Observatory, Princeton NJ 08544, USA}
\newauthor
$^8${\footnotesize Space Telescope Science Institute, Baltimore MD 21218, USA}
\newauthor
$^9${\footnotesize Dept. of Astronomy, U. Massachusetts, Amherst MA 01003, USA}
\newauthor
$^{10}${\footnotesize Dept. of Astronomy, U. Arizona, Tucson AZ 85721, USA}
}

\begin{document}

\newcommand{\lya}{{\rm Ly}$\alpha$}
\newcommand{\Lya}{{\rm Ly}$\alpha$}
\newcommand{\lyb}{{\rm Ly}$\beta$}
\newcommand{\Lyb}{{\rm Ly}$\beta$}

\newcommand{\NHI}{\mbox{$N_{HI}$}}
\newcommand{\nhi}{\mbox{$N_{HI}$}}

\newcommand{\hi}{\mbox{H~{\sc I}}}
\newcommand{\HI}{\mbox{H~{\sc I}}}
\newcommand{\HeII}{\mbox{He~{\sc II}}}
\newcommand{\CI}{\mbox{C~{\sc I}}}
\newcommand{\CIstar}{\mbox{C~{\sc I$^*$}}}
\newcommand{\CII}{\mbox{C~{\sc II}}}
\newcommand{\CIIstar}{\mbox{C~{\sc II$^*$}}}
\newcommand{\CIII}{\mbox{C~{\sc III}}}
\newcommand{\CIV}{\mbox{C~{\sc IV}}}
\newcommand{\civ}{\mbox{C~{\sc IV}}}
\newcommand{\MgII}{\mbox{Mg~{\sc II}}}
\newcommand{\Mgii}{\mbox{Mg~{\sc II}}}
\newcommand{\MnII}{\mbox{Mn~{\sc II}}}
\newcommand{\NI}{\mbox{N~{\sc I}}}
\newcommand{\NII}{\mbox{N~{\sc II}}}
\newcommand{\NV}{\mbox{N~{\sc V}}}
\newcommand{\NiII}{\mbox{Ni~{\sc II}}}
\newcommand{\OI}{\mbox{O~{\sc I}}}
\newcommand{\OIstar}{\mbox{O~{\sc I$^*$}}}
\newcommand{\OIV}{\mbox{O~{\sc IV}}}
\newcommand{\OV}{\mbox{O~{\sc V}}}
\newcommand{\OVI}{\mbox{O~{\sc VI}}}
\newcommand{\ovi}{\mbox{O~{\sc VI}}}
\newcommand{\SiII}{\mbox{Si~{\sc II}}}
\newcommand{\SiIIstar}{\mbox{Si~{\sc II$^*$}}}
\newcommand{\SiIII}{\mbox{Si~{\sc III}}}
\newcommand{\SiIV}{\mbox{Si~{\sc IV}}}
\newcommand{\AlII}{\mbox{Al~{\sc II}}}
\newcommand{\FeI}{\mbox{Fe~{\sc I}}}
\newcommand{\FeII}{\mbox{Fe~{\sc II}}}
\newcommand{\FeIII}{\mbox{Fe~{\sc III}}}
\newcommand{\Htwo}{\mbox{H$_2$}}
\newcommand{\HST}{\mbox{\it HST}}
\newcommand{\FUSE}{\mbox{\it FUSE}}
\newcommand{\ecs}{\mbox{$\rm erg~cm^{-2} sec^{-1}$}}

\newcommand{\simlt}{{\la}}
\newcommand{\simgt}{{\ga}}
\newcommand{\kms}{km~s$^{-1}$}
\newcommand{\cmsq}{cm$^{-2}$}

\maketitle

\begin{abstract}

We present an analysis of the \Lya\ forest toward 3C~273 from the Space
Telescope Imaging Spectrograph at $\sim 7$~\kms\ resolution, along  with re-processed
data from the {\it Far Ultraviolet Spectroscopic Explorer\/}.  The high UV flux of
3C~273 allows us to probe the weak, low $z$ absorbers. The  main sample consists of
21 \HI\ absorbers that we could discriminate to a sensitivity of $\log \NHI \approx 
12.5$. The redshift density for absorbers with $13.1<\log \NHI<14.0$ is $\sim
1.5\sigma$   below the mean for other lines of sight; for  $\log \NHI \geq 12.5$, 
it  is consistent with  numerical model predictions. The Doppler parameter
distribution is consistent with other low $z$ samples.      We find no evidence for a
break in the column density power-law distribution to $\log \NHI=12.3$. A broad \Lya\
absorber (BLA) is within $\Delta v\leq 50$~\kms\ and 1.3 local frame Mpc of two $\sim
0.5L^*$  galaxies, with  an \OVI\ absorber $\sim 700$~\kms\ away, similarly close to
three  galaxies and indicating overdense environments. We detect clustering on the
$\Delta v<1000$~\kms\ scale at $3.4\sigma$ significance for $\log \NHI\geq 12.6$, 
consistent with the level  predicted from hydrodynamical simulations, and indication
for a \Lya\ forest void at $0.09<z<0.12$.   We find at least two components for the
$z=0.0053$ Virgo absorber, but the total \NHI\ column is not significantly changed.

\end{abstract}


\section{Introduction}

The low redshift \Lya\ forest offers both a challenge and an opportunity to
study the physics of the intergalactic medium and its role in galaxy 
formation and evolution.  The challenge is the requirement for space-based
UV spectroscopy,  which is undergoing a renewal with the installation of COS
on \HST . The  opportunity is to observe absorption systems at small enough
distances to be able to relate nearby galaxy characteristics far down the
luminosity function.  The calculation  and interpretation of numerical
models using the fluctuating Gunn-Peterson approximation
\citep[e.g][]{Croft98} greatly  benefit from  constraints on the end epoch
for absorber distribution functions such as number densities,  column
densities, Doppler parameters, clustering and relations to galaxies.  A
consequence of the  models is that the low-$z$ \Lya\ forest is predicted to
harbour a significant fraction of the baryons, as is a hot  shocked component
\citep[e.g.][]{Dave99,Cen99,Dave01}. For a given column density, the low-$z$
forest is predicted to probe larger mass overdensities than at high
redshift, which adds complexity to comparisons of  absorber property
distribution functions between widely differing epochs.  

Exploration of the $z<1.6$ \Lya\ forest began in earnest with \HST\
observations  of the brightest QSO, 3C~273, using the Faint Object
Spectrograph \citep{Bahcall91} and Goddard High Resolution Spectrograph
\citep{Morris91}, which revealed 5 and 10  \Lya\ absorbers, respectively. 
The main result was that there were many more low redshift systems than
expected from a simple extrapolation from  $z>2$ of the decline of the
absorber redshift density over time. Concurrently, observations of the 
galaxy environments around \Lya\ absorbers were made, again with the 3C~273
field among  the first to be probed, which revealed a complex relationship
between absorbers and galaxies.  The absorber-galaxy correlation was found
to be stronger than  random but not as strong as the galaxy-galaxy
correlation  \citep{Salzer92,Morris93,Stocke95, Lanzetta95,Shull96,Tripp98},
a trend which has held up in recent years
\citep[e.g.][]{Chen05,Prochaska06}.  Using \HI -selected galaxies, however, 
\citet{Ryan-Weber06} found the absorber-galaxy correlation even stronger
than  the galaxy autocorrelation on $\sim 1-15$~Mpc scales.

Over the last 17 years, great efforts were made to accumulate observations
of a number of low redshift QSOs. A leap in observing efficiency came with
the installation of STIS in 1997.  Crucial access to higher order Lyman
lines was provided by the launch of the {\it Far Ultraviolet Spectroscopic
Explorer\/}  (\FUSE ) in 1999.   To date, a few hundred low-$z$ \Lya\ forest
lines  have been observed at resolutions down to $\sim 7$~\kms .  Such works
typically involved
studies of one to a few sight lines at a time e.g. \citet{Williger06} (Paper~I), 
but more recently therein large
analyses of archival data, including recent works by 
\citet{Lehner07a},\citet{Danforth08} ,
and  \citet{Wakker09}. The \HI\ column density distribution shows evidence  of a slight
steepening at low $z$ compared to that at $z\simgt 2$ \citep[e.g.][and
references therein]{Lehner07a}, and also a small increase in the mean Doppler
parameter $\langle b\rangle$ over time.  An excess of broad \Lya\ absorbers 
(BLAs) with $b>40$~\kms\ at $z\simlt  0.5$  is also observed, which is
attributed to a larger fraction of  warm-hot intergalactic medium (WHIM) gas
at low $z$, consistent with model predictions; this population of
BLAs is currently the
subject of active study
\citep*[e.g.][]{Richter06a,Richter06b,Tripp08,Wakker09}.    Recent
observational estimates for the baryon contribution of the low-$z$ \Lya\
forest are 30\% or  more \citep*{Penton04,Lehner07a}. \citet{Danforth08} argue
that low-$z$ \OVI\ absorbers reveal 10\% of the baryons in warm-hot gas,
but  \citet{Tripp08} have shown that many of the \OVI\ lines are
well-aligned with narrow \HI\ lines, and in those cases, the \OVI\ absorber
properties suggest an origin in cool, photoionised gas.  (A similar
conclusion is favored by \citeauthor{Thom08} 2008). Indeed, \citet{Oppenheimer09} have
argued that most low-$z$ \OVI\ systems are photoionised based on their
hydrodynamic simulations of large-scale structures.  However, some \OVI\
absorbers show strong evidence of hot gas 
\citep[e.g.][]{Tripp01,Savage05,Narayanan09}. Larger statistical studies
employing more robust diagnostics are required to reliably interpret the
natures of \OVI\ absorbers.

3C~273 has always been one of the first extragalactic targets observed with
new  technology for a variety of astronomical topics \citep[e.g.][]{Ulrich80}.   In addition to
3C~273 being extremely bright, it also lies behind the Virgo  Cluster, which
makes it particularly useful for studies of  absorber-galaxy relations and
the physical conditions in halo and filament gas,  from the cluster out to
the vicinity of the QSO. In a partial listing of successor projects after
the early \HST\ papers cited above, the sight line was the subject of
further  GHRS observations \citep{Weymann95} and a number of galaxy studies
relating to absorber  environments
\citep[e.g.][]{Salpeter95,Hoffman98,Grogin98,Impey99}.  \citet{Sembach01} 
explored the \Lyb\ forest of 3C~273 with \FUSE, and \citet{Tripp02} examined
the physical states of two  high column density Virgo Cluster absorbers. 
\citet{Rosenberg03} compared Virgo metal  systems toward RXJ1230.8+0115,
which is the nearest bright low redshift QSO  ($z=0.117$,  55~arcmin away), 
and concluded that a large scale structure filament may produce correlated 
absorption.

Preliminary results for $\sim 7$~\kms\ resolution  STIS echelle spectroscopy
of 3C~273 were presented by \citet{Heap02} and \citet{Heap03}.  We present a
full analysis here, including a  coverage of Galactic and IGM absorption
system parameters.  We also make use of  complementary data from \FUSE, with
recent improvements in data processing, reduction and analysis.    The QSO
3C~273 lies at Galactic $\ell=289.95^\circ$, $b=64.36^\circ$, and thus
serves as a very useful probe of  the Galactic halo. Over wavelengths
covered by the Ly$\alpha$ forest,  our data have $S/N=20-30$  per   $\sim
3.2$~\kms\ pixel with $1.7$ pixels per resolution element at 1200~\AA\
\citep{Dressel07}, except for small  dips at the ends of orders. The
superior signal and resolution of our data  permit us to probe the low-$z$
\Lya\ forest to $\log \NHI \sim 12.3-12.5$, which is the  best  observation
done to date, regardless of resolution.   This column density limit
corresponds to mass overdensities of $\log(\rho_H/\bar{\rho}_H)\sim 0.4\pm
0.2$ \citep{Dave99,DaveTripp01} as  calculated from hydrodynamical models in
the cold dark matter scenario.  Similar mass  overdensities at $z\sim 3$
correspond to $\log \NHI \sim 14.0-14.5$ (though the true density-column
density relation is  likely  complex due to the non-linear growth of large
perturbations).   The high $S/N$ of our spectrum also allows us to detect
BLAs with good  efficiency.   The study of hot halo gas for the Virgo
absorbers along this sight line  \citep{Tripp02,Tripp08} prompted us to
verify the component structure  of the $z=0.0053$ absorber with the \FUSE\
data. While a number of recent \Lya\ forest publications tend to include
more and more sight lines per paper, our focus on the single sight line of
3C~273 merits an in-depth analysis of the \Lya\ forest from a special
perspective, due to the high quality of the spectrum and the value added by
existing and future deep galaxy surveys in  the region.

This paper is organized as follows.  We describe the STIS and \FUSE\
observations  in \S~\ref{sec:observations}.  The absorption line selection
procedure details are  in \S~\ref{sec:absorbersample}, as are a list of
Galactic absorption lines in the  STIS data and an upper limit on \CIV\
absorption for a high \HI\ column density absorber  at $z=0.0665$. We
explain our simulations to constrain the  probability of feature detection
in \S~\ref{sec:simulatedspectra}.  Our results and discussion  for the \Lya\
forest, BLAs and their relation to \OVI\ absorbers and galaxies,  clustering
and voids, and the $z=0.0053$ Virgo absorber velocity structure are in
\S\ref{sec:resultsanddiscussion}. Our conclusions are summarised in
\S~\ref{sec:conclusions}.   We adopt a cosmology of $H_0=70$ \kms\
Mpc$^{-1}$, $\Omega_{\rm b}=0.3$ and $\Lambda=0.7$ throughout  this work.

\section{Observations and reductions}
\label{sec:observations}

\subsection{STIS spectra}

3C~273 was observed with \HST\ and STIS  using guaranteed time from Program
8017  to the STIS Instrument Definition Team (IDT). The instrumental setup used
grating E140M for seven orbits (18671 sec)  on 2000  May 2 and 2000 June 21-22,
with the   $0.2''\times 0.2''$ slit.  For the E140M grating, the STIS Instrument
Handbook gives a dispersion of $\lambda$/91700~\AA\ pix$^{-1}$, a FWHM for the 
line spread function for the $0.2''\times 0.2''$ slit
of 1.4 and 1.3 pixels at 1200 and 1500~\AA\ respectively, and a resolving power 
for a two-pixel resolution element of 45800.

The data were processed and spectra extracted with  {\sc calstis} and a
suite of programs from the STIS ID Team at Goddard Space Flight Center,
including corrections for scattered light and hot pixels and customised
merging of echelle orders to minimize echelle ripple (written by Don
Lindler).   We constructed a continuum using a combination of routines from 
automated {\sc autovp} \citep{DaveDubinski97} and interactive {\sc
line\_norm} routines  (D.~Lindler),  because the regions around emission
lines were best done with manual fitting.  Longward of our \Lya\ forest
sample region, two regions at 1402-1405 and 1425-1430~\AA\ were lost due to
vignetting from the STIS repeller wire. The total STIS wavelength coverage
is 1142--1709~\AA . Redward of 1616~\AA , there are five small
($0.15-0.58$~\AA ) inter-order gaps.

\subsection{FUSE spectra}

We use \FUSE\ spectra from the principal  investigator program P10135.  An
analysis of the \FUSE\ data was done by \citet{Sembach01}, which employed a
customised version of the \FUSE\ pipeline {\sc Calfuse}~1.8.7. In this study, we
used similar techniques and  {\sc Calfuse}~3.1.3.  We screened the data for
valid photon events, registered the individual spectra by cross-correlating
on the interstellar features, and set the zero point by cross referencing
with HST data for interstellar lines. Both the \FUSE\ LiF and SiC channel 
data show some improvement in terms  of resolution.  In the 2001 reduction,
the screening limits for pulse height distribution were (manually) 
set with care, but
in this work, the wider default values were used.  The use of these wider
values was able to reduce
the noise compared to the earlier reduction.  Aside from a small stretch in
the LiF2 data and cases where a feature fell on a fibre bundle boundary, the
velocity scale is generally good to $\pm 5$~\kms , save for a few cases
where there is  an offset of as much as half a resolution element
(22-25~\kms ).   The SiC backgrounds are good except for SiC1 at $\lambda <
920$~\AA .  We refer the reader to \citeauthor{Sembach01} for further
details.

For profile-fitting, we originally used night-only \FUSE\ data; later we tried
using all of the data but found no significant improvement in the results.  
Continua for the \FUSE\ spectra were created using interactive {\sc
line\_norm} routines.  We permitted velocity offsets as a free parameter 
on the order of
the \FUSE\ uncertainties as needed to improve individual fits.
Absorption system parameters presented in this paper use the night-only
data.

\section{Absorption Lines}
\label{sec:absorbersample}
\subsection{Selection}\label{sec:selection}

Absorption line selection was a multi-step process.  
We began the process by pre-selecting an inclusive line list  using a series
of selection criteria based on significance.   Our initial step was to make 
a preliminary selection of  absorption features from the summed STIS data
down to the $4.0 \sigma$ significance level with a Gaussian filter based on
the {\sc autovp} routine,  with half- widths of 8, 12, 16, and 20 pixels.
The large width provides sensitivity 
to broad \Lya\ absorbers and partially resolved
complexes. 
As a further preliminary (and complementary) 
constraint, we then only included significant
features within the regions flagged as significant in absorption with a
simple equivalent width significance criterion, based on a $\sim 3.3\sigma$ 
threshold for contiguous pixels below the continuum; such a criterion is
more sensitive to narrow absorbers.
Whenever possible, we
confirmed spectral features by using the \FUSE\ data.  In particular,
all (9 out of 9) absorbers with $\log \NHI \geq 13.24$ over $0.003<z<0.147$
were confirmed by inspection of \Lyb\ data, and \Lyb\ inspections showed no 
useful information for lower column density \Lya\ forest systems. The
selection algorithm to this point was designed to be largely  objective and
repeatable, aside from continuum fitting around emission features.   We then
visually checked each absorption feature to avoid including noise spikes or potential
continuum errors.

We subsequently tested the  effectiveness of this selection technique by
subjecting it to simulated spectra with similar resolution to the data to
determine the 80\% detection probability based on a grid of Doppler
parameters, \HI\ column densities and $S/N$ values.  Details 
are in \S\ref{subsec:simulatedspectra} below. The
decrease in detection probability drops steeply from nearly 100\% to nearly 
zero  over 0.1 dex in column density for a given $S/N$ and Doppler
parameter, which enables straightforward determinations of the 80\%
detection efficiency to  $\simlt $0.05  dex accuracy.  Such a steep decline
is not unreasonable given that the absorbers are of such low
column density ($\log \NHI \sim 12.5$) that they are well on the linear
part of the curve of growth.  Their equivalent widths change little over
the limited range of Doppler parameters and S/N (factors of 2) 
which we explored, and their velocity widths also only vary over a small 
interval.
Therefore, when detecting weak but significant features relying in part on
a boxcar approximation to an absorption profile, a range of 0.3 dex 
(factor of 2) in column density  would be a large range to expect a
uniform detection threshold probability.

We made a concentrated effort to understand the rate of false detections,
due to our desire to probe as deeply as possible down the \hi\ column
density distribution. Therefore, as an additional test, we examined the
spectrum $>5000$~\kms\ redward of the QSO's \Lya\ emission all the way to
the red end of the STIS data (1710~\AA ) for unidentified absorption
features to test for the frequency of appearance of potential  spurious
features.   The most significant one at 1667.9~\AA\  had a significance
$2.3\sigma$, which was well below our preliminary selection threshold.  It
would have  \Lya\ fitting parameters of $\log \NHI = 12.53\pm 0.13$ and
Doppler parameter $b=13.4\pm 6.3$.  The  relatively  large (47\%) error in
$b$ would make it unlikely to be considered a reliable \Lya\ absorber, even
if it had somehow been included in our preliminary absorber list. The
feature does not correspond to any plausible metal line for any \Lya\
absorber in our sample. The final STIS E140M spectrum of 3C~273, with
absorption systems marked,  is shown in Figure~\ref{fig:plotspec}.

\subsection{Profile fitting technique}\label{sec:fitting}

The data were profile-fitted with  {\sc
vpfit}\footnote{http://www.ast.cam.ac.uk/$\sim$rfc/vpfit.html}
\citep{Webb87}  using Voigt profiles convolved  with the local STIS line
spread function (LSF) taken from the STIS Instrument Handbook.   Oscillator
strengths are from the {\sc vpfit} atomic data list (2004 Dec. version),
based on \cite{Morton03}. Multiple transitions were used for simultaneous
fits whenever they could improve constraints on column densities and Doppler
parameters. We also used, as necessary,  the {\sc vpfit} capabilities to
allow for offsets in the local continuum level and in velocity for each
spectral interval used in a fit.  The latter would in particular compensate
for small velocity uncertainties in the \FUSE\ data.

\clearpage
\begin{figure}
\includegraphics[width=175mm]{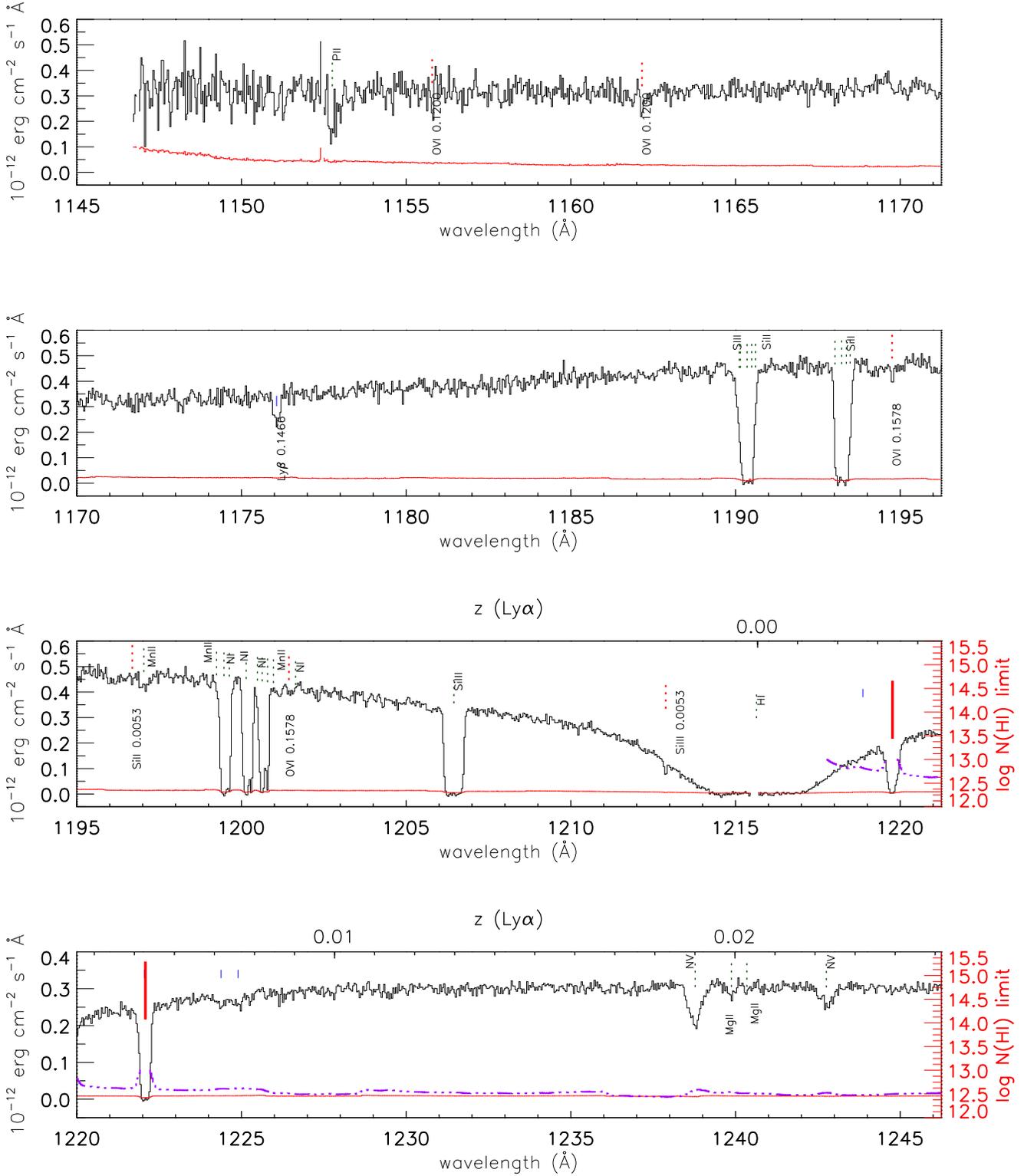}
\caption[3C~273 STIS spectrum]{3C~273
data and $1\sigma$ errors {\it (red)} binned
by 3 pixels for presentation only. 
{\it Grey dash-triple dotted curve:} 80\% detection threshold in $\log N(\HI)$
(right axis).
{\it Long, bold (red) ticks:} \Lya\ lines with metals. 
{\it Long unlabelled (blue) ticks:} \Lya\ forest ($z$ on upper axis)
with $\log \NHI< 12.5$ in the complete
survey (\S~\ref{sec:observations}). {\it Thick (blue) tick:} reliable BLA.
{\it Short (blue) ticks:} other \Lya\ lines and
higher order Lyman lines (labelled).
{\it Heavy dotted (red) ticks:} intervening metal lines (labelled).
{\it Medium dotted (green) ticks:} Galactic metal absorption.
\label{fig:plotspec}
}
\end{figure}

\clearpage
\setcounter{figure}{0}
\begin{figure}
%\epsscale{0.9}
\includegraphics[width=185mm]{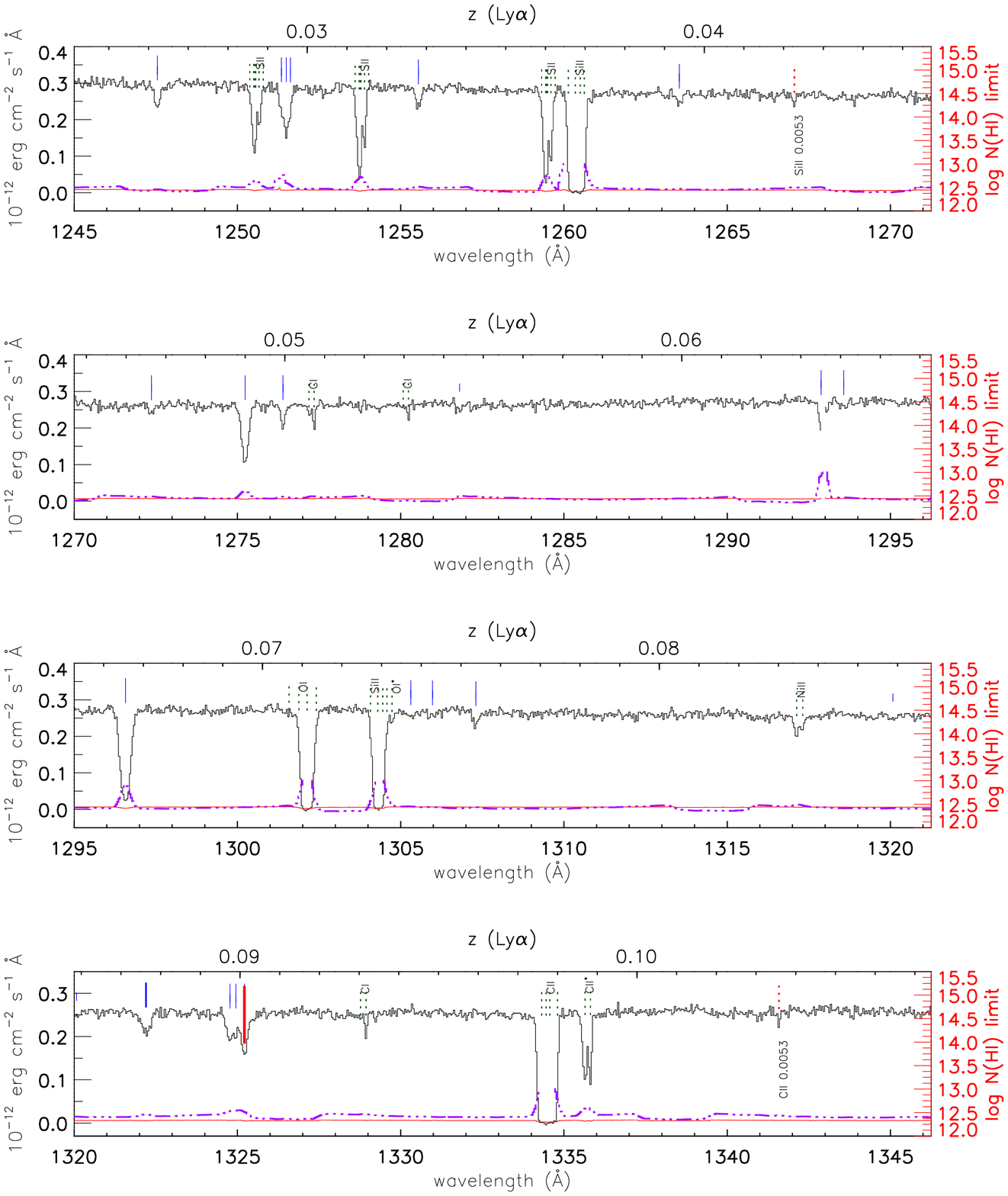}
\caption[3C~273 STIS spectrum]{
Continued.
}
\end{figure}

\clearpage
\setcounter{figure}{0}
\begin{figure}
%\epsscale{0.9}
\includegraphics[width=185mm]{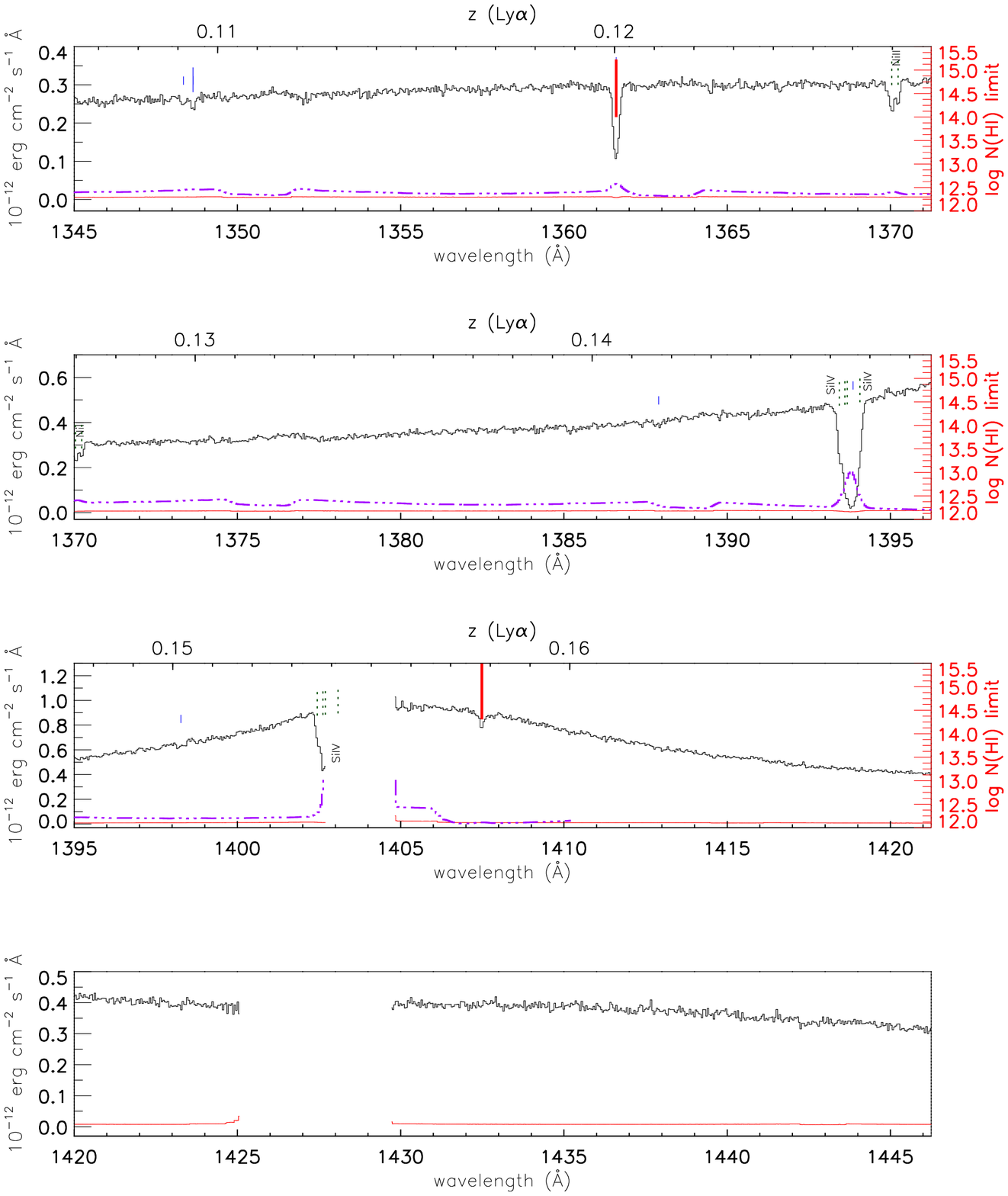}
\caption[3C~273 STIS spectrum]{
Continued.
}
\end{figure}

\clearpage
\setcounter{figure}{0}
\begin{figure}
%\epsscale{0.9}
\includegraphics[width=185mm]{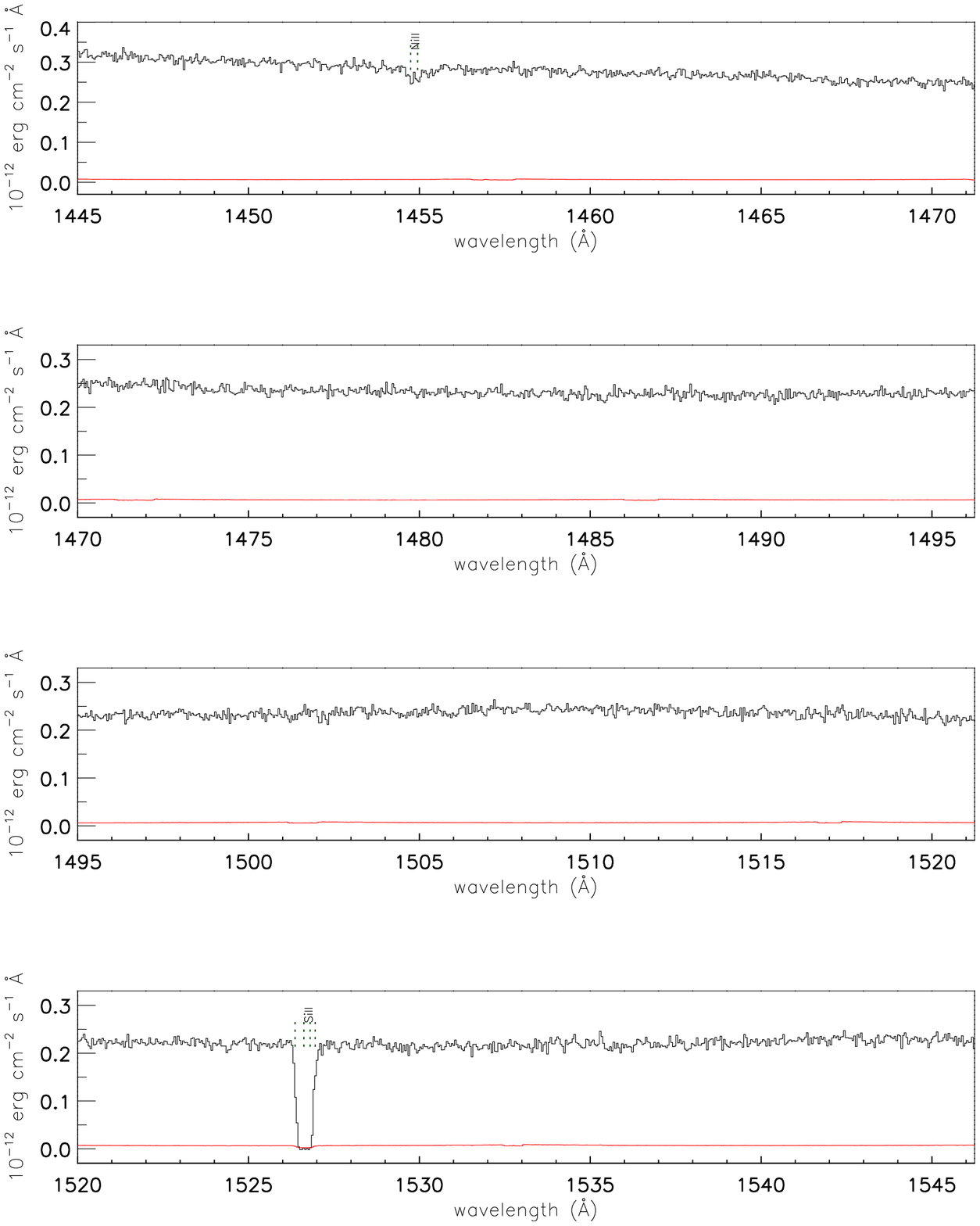}
\caption[3C~273 STIS spectrum]{
Continued.
}
\end{figure}

\clearpage
\setcounter{figure}{0}
\begin{figure}
%\epsscale{0.9}
\includegraphics[width=185mm]{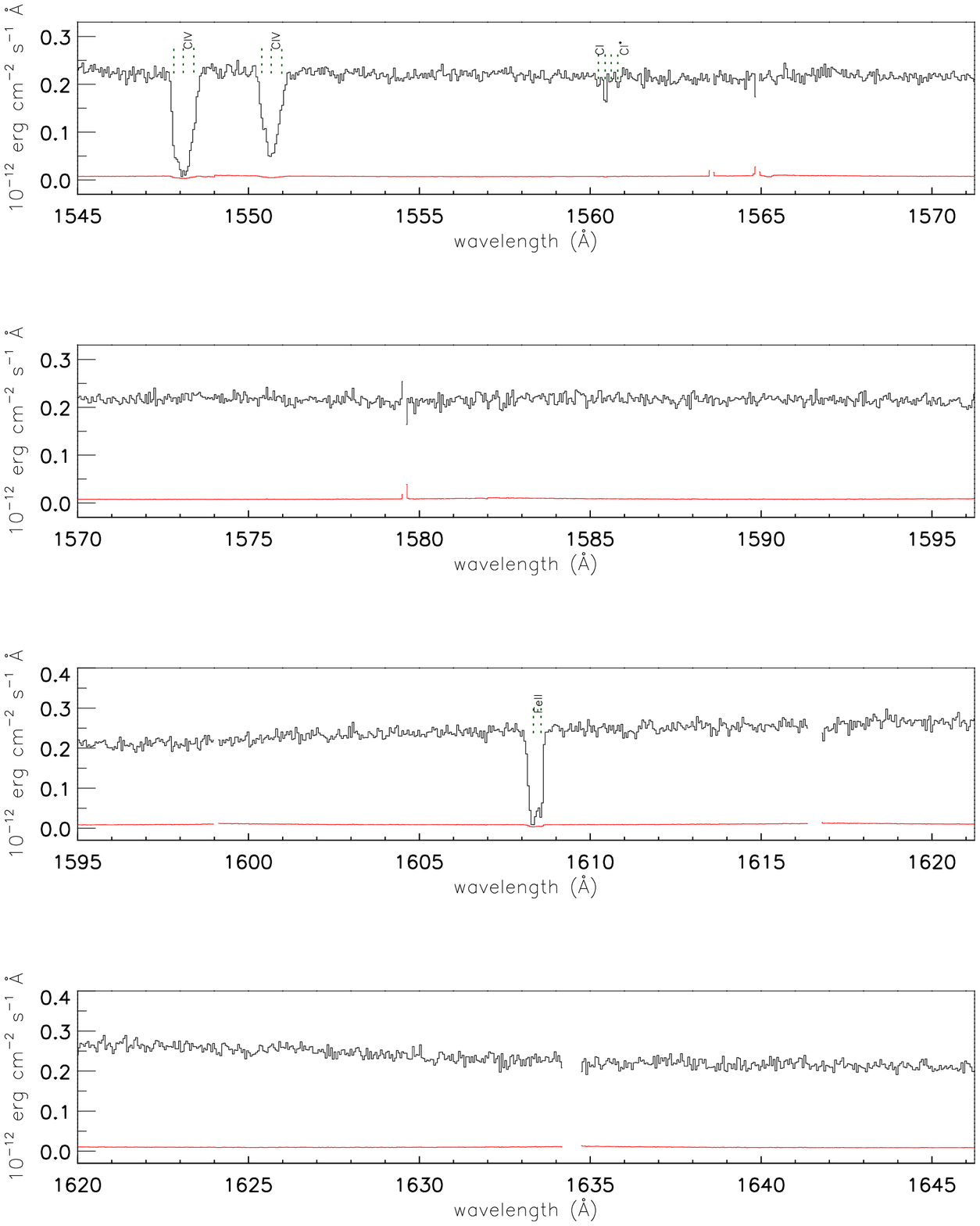}
\caption[3C~273 STIS spectrum]{
Continued.
}
\end{figure}

\clearpage
\setcounter{figure}{0}
\begin{figure}
%\epsscale{0.9}
\includegraphics[width=185mm]{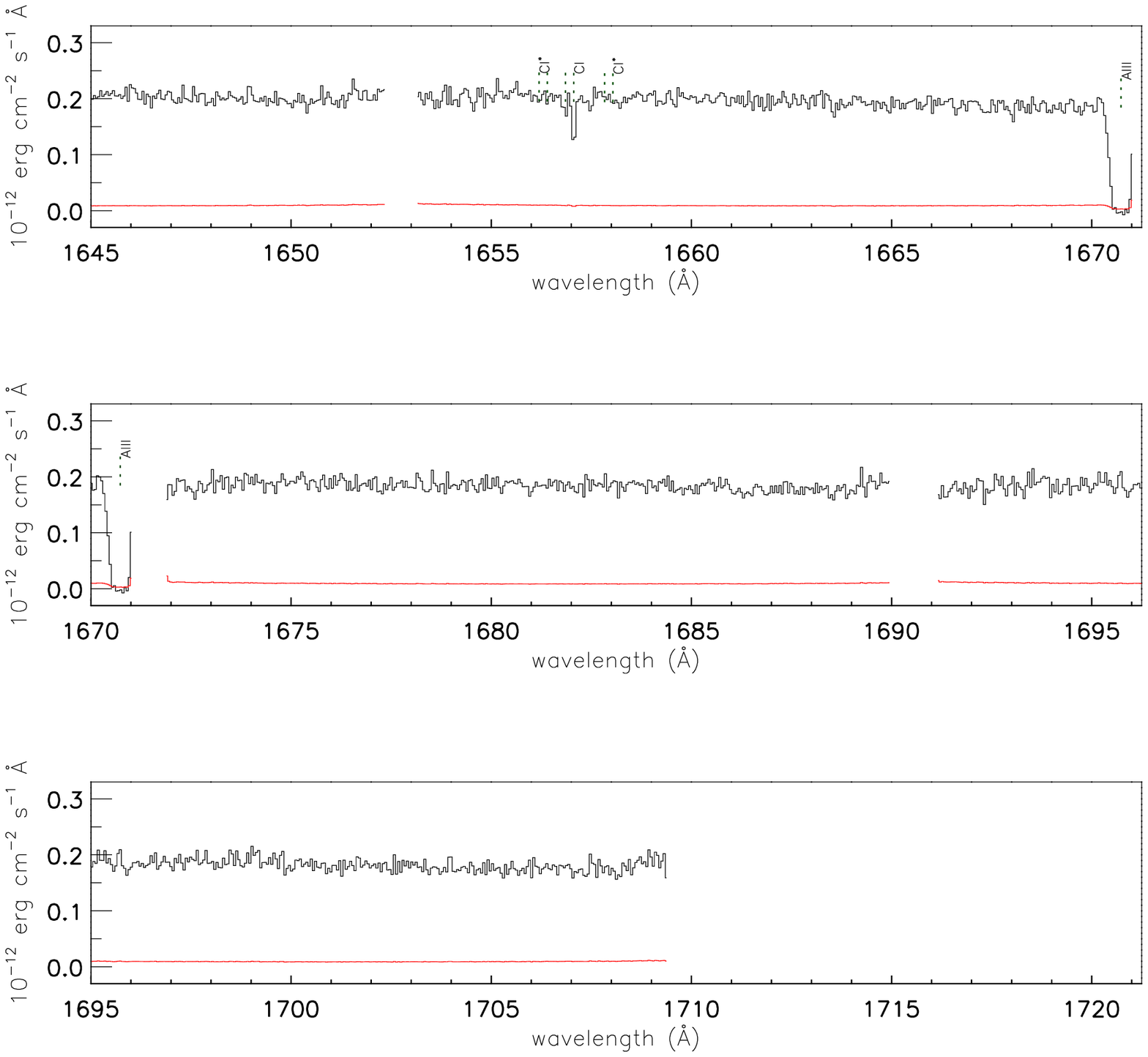}
\caption[3C~273 STIS spectrum]{
Continued.
}
\end{figure}
\clearpage

\subsection{Simulated spectra to determine detection probabilities}
\label{subsec:simulatedspectra}

To give us a clearer picture of our \Lya\ line detection probability as a 
function of S/N ratio and Doppler parameter, we analysed 880 simulated
Ly$\alpha$ lines. The emphasis here is on characterizing the detection  and
recovery of of weak absorbers and their line parameters,  because the 3C~273
data uniquely probe the low end of the  \Lya\ forest column density
distribution. The simulations were in addition to the ones done for
PKS~0405-123 described in Paper~I.  The simulated lines were generated using
the STIS LSF and a grid of values for $S/N$, Doppler parameter and \HI\
column density. The $S/N$ values were set to either 20 or 40 per pixel,
Doppler parameters were set to ranges with averages of 17.5, 22.2 or 24.7
and 35.0~\kms, and the input log~\nhi\  values varied over  12.21--12.80 for
S/N=20 and 11.85--12.55 for $S/N=40$. The simulated spectra were
continuum-normalised, so any errors, systematic or random, involving
continuum fitting are not reflected in the subsequent  simulation analysis.
There were also no noise spikes inserted into the spectra. We calculated
statistics comparing the line parameters  for simulated lines which were
successfully recovered {\it vs.} the input values.  Results for the various
combinations of input S/N ratio and Doppler parameter are shown in
Tables~\ref{tab:simulations_b} and \ref{tab:simulations_nhi}. The mean
recovered measured Doppler parameters and column densities match the mean
input values within the mean of the profile fitting errors.  The marginal
tendency for the narrowest lines to be recovered with higher Doppler
parameters is likely due to the effects of noise broadening input
for features near the detection limit for recovered lines; noise which
would make lines narrower presumably would cause the features to drop
below the (essentially equivalent width-defined) 
detection threshold \citep{Rauch93}.
The recovery
of the line parameters (e.g., the $b$-value) was found to be even better for
stronger lines detected at high significance, up until the point where the
\HI\ lines start to saturate badly. The spurious detection rate is $<1$\%
among all the simulations above the 80\% detection probability threshold,
and thus is not considered significant. The boundary of the 80\% detection
probability threshold  can be parametrically fitted by $\log \NHI = 12.11
+0.27  \log (b/24.7) - 1.12 \log ({\rm snr}/40.)$, where snr is the signal
to noise
ratio (S/N) per pixel, and established using simulated lines with $16\simlt
b\simlt  36$~\kms .
The process was adapted from the method used in Paper~I.
\footnote{The coefficient of 0.27 for $\log (b/24.7)$ for
this paper  is $\sim 25$\% lower than the analogous value of 0.34 from
Paper~I, and the {\it snr} coefficient in this work is 11\% higher.  The
difference in the $\log (b/24.7)$ coefficient may be caused by the
relatively lower column density range used for the input lines for this
study, in that the limited Doppler parameter range used affects the
equivalent widths and thus detectibility of the lines less strongly.   The
difference in the {\it snr} coefficient may also stem from a similar cause,
or from either the higher S/N in the 3C~273 data and/or the use of only two input S/N
values in this study, 20 and 40.  In comparison, for PKS~0405-123,  we used
three evenly  distributed in log space over the range we tested. The
difference in results between a coefficient of 1.12 and 1.00 (without
adjusting the other coefficients)  is only 0.03 dex for our minimum S/N of
21, and the current value of 1.12 leads to a more conservative detection
threshold as $snr/40<1$.  We do not believe that  the results of this paper
are significantly affected by the difference in parametrization.} \footnote{We will
publish an erratum for Paper~I (Williger et al., in  prep.).  In that work, a
subset of  absorbers principally below the 80\% detection threshold  had
erroneously high calculated significance because the contribution of
continuum uncertainties to the total equivalent width errors
was not included in the software.}
We consider this parametrization useful to determine our 80\% detection threshold
over a Doppler parameter range of 10-40~\kms , with the bounds
based both on our simulations and to compare with the 
analysis of \citet{Lehner07a}.  

\clearpage
\clearpage
\begin{table}
%\tablewidth{0pc}
\caption{Statistics of simulated data: Doppler parameters
\label{tab:simulations_b}}
%\tablehead{ 
%\colhead{$S/N$\tablenotemark{a}} & \multicolumn{2}{c}{Doppler parameter (\kms )} & \multicolumn{2}{c}{Doppler parameter (\kms )} & \multicolumn{2}{c}{Doppler parameter (\kms )} \\
%\colhead{}    & \colhead{input\tablenotemark{b}} & \colhead{recovered\tablenotemark{c}} & \colhead{input\tablenotemark{b}} 
%& \colhead{recovered\tablenotemark{c}} & \colhead{input\tablenotemark{b}} & \colhead{recovered\tablenotemark{c}}    }
%\startdata  
\begin{tabular}{ccccccc} \hline
$S/N^a$       & \multicolumn{2}{c}{Doppler parameter (\kms )} & \multicolumn{2}{c}{Doppler parameter (\kms )}& \multicolumn{2}{c}{Doppler parameter (\kms )}\\
              &   input$^b$             & recovered$^c$       & input$^b$       &  recovered$^c$      & input$^b$       &  recovered$^c$      \\ \hline
  20       & $17.2\leq b\leq 17.8$	& $19.0\pm 5.0(4.1)$  & $22.2\pm 5.5$	&  $21.8\pm 8.0(5.5)$ & $35.0\pm 0.6$	& $35.3\pm 8.1(8.1)$  \\
  40       & $17.2\leq b\leq 17.8$	& $19.5\pm 6.8(4.5)$  & $24.7\pm 0.1$	&  $26.7\pm 7.0(6.3)$ & $35.0\pm 0.6$	& $35.0\pm 7.0(7.2)$ \\ \hline
%\enddata
%\vspace{2mm}
\medskip
%\tablenotetext{a}{Signal to noise ratio per pixel in simulated STIS spectrum.}
%\tablenotetext{b}{Mean and first order about the mean for input Doppler parameters in simulations
%{\it which were recovered}.}
%\tablenotetext{c}{Mean and first order about the mean for recovered Doppler parameters in simulations,
%   with the mean in the formal $1\sigma$ profile fitting errors in parentheses.  The mean recovered Doppler
%   parameters are all within the mean profile fitting errors of the input values.}
\end{tabular}
$^a$ Signal to noise ratio per pixel in simulated STIS spectra.\\
$^b$ Mean and first order about the mean for input Doppler parameters in simulations {\it which were recovered}.
The simulations for S/N=20, $\langle b\rangle = 22.2$~\kms , were made from a distribution made at
the beginning of our analysis, with a much
wider standard deviation (5.5~\kms ) in input $b$ than the others, hence the difference with the
$\langle b\rangle = 24.4$~\kms\ case (the geometric mean of 17.5 and 35.0).  
The results between the two cases are within the scatter, so we
did not make a simulation run with S/N=20, $\langle b\rangle = 24.7$~\kms .   \\
$^c$ Mean and first order about the mean for recovered Doppler parameters in simulations,
   with the mean in the formal $1\sigma$ profile fitting errors in parentheses.  The mean recovered Doppler
   parameters are all within the mean profile fitting errors of the input values.\\
\end{table}

%\clearpage
\begin{table}
\caption{Statistics of simulated data: \HI\ column densities
\label{tab:simulations_nhi}}
\begin{tabular}{rccc} \hline
%\tablewidth{0pc}
%\tablehead{ 
%\colhead{$S/N$\tablenotemark{a}} & \colhead{Doppler\tablenotemark{b}}  & \multicolumn{2}{c}{$\log N(\HI )$}  \\
%\colhead{}    & \colhead{parameter (\kms )} & \colhead{input\tablenotemark{c}} & \colhead{recovered\tablenotemark{d}}   }
%\startdata  
$S/N^a$   &Doppler$^b$& \multicolumn{2}{c}{$\log N(\HI )$}     \\
          & parameter (\kms ) & input$^c$ & recovered$^d$      \\ \hline
20        & 17.5 &     $12.54\pm 0.12$ & $12.55\pm 0.12(0.07)$  \\
40        & 17.5 &     $12.17\pm 0.12$ & $12.20\pm 0.11(0.09)$  \\
&&\\
20        & 22.2 &     $12.41\pm 0.10$ & $12.41\pm 0.13(0.08)$  \\
40        & 24.7 &     $12.26\pm 0.12$ & $12.28\pm 0.13(0.08)$  \\
&&\\
20        & 35.0 &     $12.64\pm 0.12$ & $12.65\pm 0.13(0.08)$ \\
40        & 35.0 &     $12.36\pm 0.12$ & $12.37\pm 0.13(0.07)$ \\ \hline
%\enddata
%\vspace{2mm}
\medskip
\end{tabular}
$^a$ Signal to noise ratio per pixel in simulated STIS spectrum.\\
$^b$ Simulation sets are specified by input Doppler parameter.\\
$^c$ Mean and first order about the mean for input log \HI\ column densities in simulations {\it which were recovered}.\\
$^d$ Mean and first order about the mean for recovered log \HI\ column densities in simulations,
   with the mean in the formal $1\sigma$ profile fitting errors in parentheses.  \\
\end{table}

\clearpage

\subsection{Profile fits to data}
\label{sec:simulatedspectra}

In addition to \hi\ lines, we find and list,  for
completeness, profile fits or column density limits  \citep[from {\sc vpfit} or 
from the apparent
optical depth method,][]{Savage91}  for a number of Galactic and  intervening
metal absorber species.  

{\it Galactic absorption.} We find Galactic absorption from \HI , \CI ,
\CIstar, \CII ,  \CIIstar , \CIV ; \NI , \NV , \OI , \MgII , \AlII , \SiII ,
\SiIII , \SiIV , P\,{\sc II}, S\,{\sc II}, S\,{\sc III}, Mn\,{\sc II}, \FeII ,
Ni\,{\sc II}. 
Galactic absorption ranges over $-135 < v < 68$ \kms ,
consistent with the high velocity absorption found in the \FUSE\  spectrum  by
\cite{Sembach01}. A complete list of absorber profile fitting parameters  is
in Table~\ref{tab:linelist_gal}.   In the case of saturated lines, we
calculate  lower limits  by the apparent optical depth method
\cite{Savage91}, which are consistent with the results of \cite{Sembach01}.
The maximum optical depth used is a function of local S/N  and continuum
error ($\tau_{max}\sim \ln [\sqrt 2/(S/N)]$ per pixel), in which both
effects  contribute to similar degrees. For the 3C~273 data, $\tau_{max}\sim
2.3-3.0$.

For two of the highest \NHI\ column density systems toward 3C~273, members
of  this collaboration and others  have investigated metallicity limits and
undertaken studies in the low $z$ IGM 
\citep[e.g.][]{Dave01,Tripp02,Stocke07,Tripp08}.   Such low redshift, high
column density absorbers are extremely useful for comparison with nearby low
luminosity faint galaxies e.g. the dwarf galaxy found 71~kpc away from the
$z=0.0053$ Virgo cluster absorber \citep{Stocke04}. For completeness and 
comparison with their results, we provide here upper limits for \CIV\ for each of
these \NHI\ absorbers, and for a sum of the \CIV\ regions over a consistent 
$-100\leq \Delta  v\leq 100$~\kms , using the above-mentioned apparent
optical depth method. The $z=0.003369$ absorber was found by
\citeauthor{Tripp02} to have rest equivalent width  $W_{r,CIV}<27$~m\AA\
over $-100<\Delta v<100$~\kms , or $\log N(\CIV)<12.8$ assuming an absorber 
on the linear part of the curve of growth.   \citeauthor{Stocke07} found
$\log N(\CIV)<12.27$ (3$\sigma$) for the $z=0.066548$ system over
$-50<\Delta v<50$~\kms . We measured an upper limit for the $z=0.066548$
absorber of $W_{r,CIV}\simlt 50$~m\AA\ ($3\sigma$) over $-100<\Delta
v<100$~\kms , or $\log N(\CIV)\simlt 13.1$.\footnote{The difference in upper
limits arises in that the \citeauthor{Stocke07} limit is for one resolution
element, whereas our limit is over 200~\kms\ in velocity space or $\sim 28$
resolution elements, to be consistent with the \citeauthor{Tripp02}
measurement.} We stacked the data for the two systems and found 
$W_{r,CIV}\simlt 19$~m\AA\ ($3\sigma$)  $-100<\Delta v<100$~\kms , or $\log
N(\CIV)\simlt 12.7$.\footnote{This limit is slightly tighter than the
24~m\AA\ result expected from stacking the spectra according to the inverse
square of the S/N.  The difference in our result may be
from the way each team determined the local continuum
and S/N, in particular as these were measured over two different
velocity intervals.}

\section{Results and discussion}
\label{sec:resultsanddiscussion}
\subsection{Ly$\alpha$ absorber statistics}
\label{subsec:lyaforeststats}

We use our detection probability results to define a sample for $\log (N(\HI
))\geq 12.5$,  which is approximately the 80\% detection threshold for our
worst case $S/N=20$ in the \Lya\ forest and assuming a Doppler parameter of
$b=40$~\kms\ \footnote{ Note that by choosing $b=40$~\kms , this leads to a
higher, more  conservative 80\% probability detection threshold in \NHI\ by
0.06 dex than by choosing  $b=25$~\kms , which is typical of high
resolution, high $S/N$ \Lya\ forest data.  We prefer to be conservative,
given the possibility of unresolved blends caused by our  limits in the
achievable $S/N$.  This choice of $b$ does not affect our subsequent
analysis in any way except for this small change  in the threshold for $\log
\NHI$.}.   This low \HI\ column density threshold sample covers the range
$0.020<z<0.139$ and contains  21 \Lya\ absorbers. We excluded absorbers
within $\Delta v\sim  5000$~\kms\ of 3C~273 to minimize the impact of the
proximity effect \citep{Murdoch86,Bajtlik88}.

In later subsections, we will discuss in detail our examinations of the 
Ly$\alpha$ forest Doppler parameter distribution, redshift density
$dN/dz$, broad line population and column density distribution. We do not
distinguish between metal and Ly$\alpha$-only systems in this analysis,
the same approach as in 
Paper~I. Metals have  been found in progressively lower column
density \Lya\ systems at $z\simgt 2$  \citep[e.g.][and references
therein]{Cowie95, Songaila96, Pettini01}.  Also, the absorbers we can
detect at low redshift probably are best related to large density
perturbations at $z\simgt 2$ \citep[e.g.][]{Schaye01} which are commonly
associated with  metals. Finally,  the limited wavelength range of the 
data does not permit a uniform coverage for metal detection (in
particular for \CIV , limited to $z\simlt  0.1$).

\clearpage
\clearpage
\begin{table}
\caption{Galactic absorption lines toward 3C~273 from STIS data
\label{tab:linelist_gal}}
%\tablehead{ 
%\colhead{species} & \colhead{$z$} & \colhead{error\tablenotemark{a}}     &\colhead{$b$ (\kms )} & \colhead{error\tablenotemark{a}} 
%& \colhead{$\log N$} & \colhead{error\tablenotemark{a,b}} 
%& \colhead{lines (\AA )/remarks\tablenotemark{c,f,g}}} 

\begin{tabular}{lccrrrcl} \hline
species & $z$ & error$^a$ & $b$ (\kms ) &  error$^a$ & $\log N$ (\kms ) &  error$^a$ & lines (\AA )$^{c,f,g}$ \\ \hline

%\tablewidth{0pc}
\footnotesize
%\startdata

%NB check for lower limits to column densities for saturated lines! GMW 14 Jan 2006
$\!\!\!\!$~O I   & -0.000451 & 0.000041 &   61.5 &   22.1 &  13.55 &   0.12 & 1302     \\
 C IV  & -0.000240 & 0.000003 &   11.6 &    1.8 &  13.45 &   0.09 & 1548,1550	   \\
 \SiIV\  & -0.000230 & 0.000003 &    7.9 &    1.4 &  12.57 &   0.06 & 1393,1402	  \\
 \SiII\  & -0.000224 & \ldots   &\ldots  &\ldots  &$>13.24$&\ldots  & 1190,1193,1260,1304,1526	 \\
 O I   & -0.000220 & \ldots   &\ldots  &\ldots  &$>13.59$&\ldots  & 1302     \\
 S II  & -0.000159 & 0.000016 &   10.1 &    4.9 &  13.74 &   0.24 & 1250,1253,1259     \\  
% C II  & -0.000158 & 0.000058 &   15.1 &   13.3 &  15.24 &   2.54 & 1334  GET LIM   \\
 \SiIV\  & -0.000103 & \ldots   &\ldots  &\ldots  &$>13.19$&\ldots  & 1393,1402	  \\
% O I*  & -0.000093 & 0.000006 &   12.9 &    2.9 &  13.41 &   0.06 & 1304 (not secure)\\ 2.8 sigma, continuum shift redward of echelle order boundary
 S III & -0.000084 & \ldots   &\ldots  &\ldots  &$>14.26$&\ldots  & 1190     \\
% \FeII\  & -0.000071 & 0.000002 &   22.6 &    0.8 &  14.76 &   0.02 & 1608     \\
 \FeII\  & -0.000071 & \ldots   &\ldots  &\ldots  &$>14.71$&\ldots  & 1608     \\
% N I   & -0.000067 & 0.000001 &   17.6 &    0.5 &  15.08 &   0.03 & 1199,$\lambda\lambda$1200	  \\
 N I   & -0.000067 & \ldots   &\ldots  &\ldots  &$>15.04$&\ldots  & 1199,$\lambda\lambda$1200	  \\
 \MnII\  & -0.000067 & \ldots   &   14.2 &\ldots  &  12.60 &   0.19 & 1197,1199,1201$^d$     \\
 \NiII\  & -0.000067 & 0.000004 &   22.3 &    1.7 &  13.73 &   0.03 & 1317,1370,1454	  \\
 C II  & -0.000067 & \ldots   &\ldots  &\ldots  &$>15.54$&\ldots  & 1334     \\
 C IV  & -0.000066 & 0.000003 &\ldots  &\ldots  &$>14.46$&\ldots  & 1548,1550 \\							 
% S II  & -0.000061 & 0.000002 &   14.7 &    0.9 &  15.13 &   0.03 & 1250,1253,1259     \\
 S II  & -0.000060 & \ldots   &\ldots  &\ldots  &$>15.22$&\ldots  & 1250,1253,1259     \\
 \SiIV\  & -0.000060 & 0.000012 &   48.8 &    2.3 &  13.73 &   0.04 & 1393,1402	  \\
 P II  & -0.000056 & 0.000017 &   29.1 &    7.5 &  13.69 &   0.10 & 1152,1532	  \\
 \SiII\  & -0.000067 & \ldots   &\ldots  &\ldots  &$>15.03$&\ldots  & 1190,1193,1260,1304,1526	 \\
 C II* & -0.000045 & 0.000002 &   14.3 &    0.8 &  13.83 &   0.02 & 1335     \\
 C I*  & -0.000044 & 0.000000 &    9.1 &    0.0 &  12.24 &   0.26 & 1561,1656,1657$^e$ \\
 C I   & -0.000044 & 0.000005 &    9.1 &    2.4 &  12.79 &   0.07 & 1277,1280,1328,1560,1656	 \\
% \SiIII\ & -0.000042 & 0.000001 &   29.9 &    0.9 &  15.69 &   0.16 & 1206     \\
 \SiIII\ & -0.000042 & \ldots   &\ldots  &\ldots  &$>14.02$&\ldots  & 1206     \\
 Mg~II  & -0.000038 & 0.000012 &   27.9 &    5.1 &  15.43 &   0.06 & 1239,1240	  \\
 N V   & -0.000037 & 0.000005 &   59.8 &    2.2 &  13.93 &   0.01 & 1238,1242	  \\
 Al~II  & -0.000033 &\ldots    &\ldots  &\ldots  &$>13.62$&\ldots  & 1670 \\
% S II  & -0.000032 & 0.000008 &    1.9 &    1.7 &  14.40 &   0.37 & 1250,1253,1259     \\
 H I   & -0.000027 & 0.000005 &\ldots  &\ldots  &  20.23 &   0.00 & Ly$\alpha,\beta$	 \\ % check with Bob, PKS0405 - is b-value 30.9+-1.2 accurate, or better \ldots ?
 O I   & -0.000025 & \ldots   &\ldots  &\ldots  &$>15.38$&\ldots  & 1302     \\
% S II  & -0.000014 & 0.000003 &    0.8 &    1.0 &  15.05 &   2.36 & 1250,1253,1259     \\
% C II  &  0.000025 & 0.000099 &   22.2 &   15.4 &  16.26 &   0.84 & 1334     \\
% \SiII\  &  0.000064 & 0.000010 &    7.2 &    2.1 &  15.22 &   0.45 & 1190,1193,1260,1304,1526	 \\
% \FeII\  &  0.000068 & 0.000002 &    9.0 &    0.8 &  14.36 &   0.04 & 1608    \\
 \FeII\  &  0.000068 & \ldots   &\ldots  &\ldots  &$>14.28$&\ldots  & 1608     \\
% N I   &  0.000068 & 0.000001 &    6.8 &    0.8 &  14.88 &   0.20 & 1199,$\lambda\lambda$1200	  \\
 N I   &  0.000068 & \ldots   &\ldots  &\ldots  &$>14.66$&\ldots  & 1199,$\lambda\lambda$1200	  \\
 \MnII\  &  0.000068 & \ldots   &    4.4 &\ldots  &  12.63 &   0.18 & 1199,$\lambda\lambda$1200$^d$	  \\
 S II  &  0.000069 & 0.000001 &    9.4 &    0.6 &  14.75 &   0.01 & 1250,1253,1259     \\
 \NiII\  &  0.000074 & 0.000004 &   13.3 &    2.0 &  13.41 &   0.05 & 1317,1370	  \\
 C II* &  0.000078 & 0.000001 &    7.6 &    0.7 &  13.73 &   0.02 & 1335     \\
 C I   &  0.000081 & 0.000001 &    6.4 &    0.7 &  13.27 &   0.02 & 1277,1280,1328,1560,1656	 \\
 C I*  &  0.000081 & 0.000000 &    6.4 &    0.0 &  12.67 &   0.09 & 1561,1656,1657$^e$  \\
 C IV  &  0.000138 & 0.000012 &   20.7 &    4.4 &  13.32 &   0.15 & 1548,1550	     \\
 \SiII\  &  0.000159 & \ldots   &\ldots  &\ldots  &$>14.59$&\ldots  & 1190,1193,1260,1304,1526	 \\
 S II  &  0.000171 & 0.000009 &    1.3 &    4.1 &  13.13 &   0.21 & 1250,1253,1259     \\
 O I   &  0.000190 & 0.000015 &   13.0 &    5.7 &  13.41 &   0.19 & 1302     \\
 C II  &  0.000206 & \ldots   &\ldots  &\ldots  &$>14.89$&\ldots  & 1334     \\
 \SiIV\  &  0.000225 & 0.000020 &   15.2 &   11.6 &  12.21 &   0.56 & 1393	  \\ \hline

%\enddata
%\vspace{2mm}
\medskip
\end{tabular}
$^a$ Errors are 1~$\sigma$, and assume that the component structure is correct.\\
$^b$ Upper limits from {\sc vpfit} or the apparent optical depth method of \citet{Savage91}.\\
$^c$ \phantom{.}$\!$Lines used in profile fits.\\
$^d$ Mn~{\sc II} tied in redshift and Doppler parameter to N~{\sc I}.\\
$^e$ \CI$^*$ tied in redshift and Doppler parameter to \CI .\\
$^f$ We find an unidentified feature at 1149.3~\AA\ (also in the FUSE data).\\
$^g$ We find a weak feature at 1304.74~\AA\ to be an artefact just redward of an echelle
overlap region, and not securely identifiable with any plausible ISM or IGM absorption.\\
%$^e}{FUSE data used for \ion{P}{2} upper limit column density.}
%$^f}{Column density from \citet{Prochaska04}.}
%$^g}{The \CIV\ fit is from archival STIS G230M data.}
\end{table}

\clearpage

\subsubsection{Redshift density and intervening absorber list}
\label{subsubsec:redshiftdensity}

Measuring the \Lya\ forest redshift density has been one of the biggest 
goals among \HST\ Key Projects, and initially yielded surprising results.
From among  the first results with FOS \cite{Bahcall91} and GHRS
\cite{Morris91} toward 3C~273, the \Lya\ redshift density was found to be
higher than expected from extrapolations of ground-based data.  To
compare results with recent high resolution data in   the literature, we
present data for $13.1\leq \log \NHI < 14.0$ and $14.0\leq \log  \NHI <
17.0$, using the restrictions of $b\leq 40$~\kms , 
and errors in $b$ and
$\NHI$ of $\leq 40$\% for both 3C~273 and the literature values
\citep[to use the same criteria as][]{Lehner07a}. We also
present results for  $\log \NHI \geq 12.5$. Due to the low redshift of
3C~273, the data are  unaffected by losses from inter-order gaps in the
STIS data.  Similar to our analysis in  Paper~I, we excluded pixels from
metal absorption  with optical depth $\tau \geq 2$, which removed $\Delta
z=0.002$ from the samples for $\log \NHI \geq 13.1$ and $\log \NHI \geq
14.0$.\footnote{Metal absorption from pixels with $\tau<2$ can also
reduce the probability of detection an IGM \Lya\ absorber, but less so
than those with $\tau\geq 2$.  The value of $\tau=2$ is a compromise to
eliminate the worst metal absorption regions without losing too many
pixels with recoverable \Lya\ data. Other factors such as multiple
transitions from a metal ion can mitigate the masking effect of metal 
absorption. For example, we recovered the presence of \Lya\ absorption
blended with Galactic  \SiIV .} This left an unblocked redshift path of
$\Delta z = 0.135$ for $\log \NHI \geq 13.1$ and  for $\log \NHI \geq
14.0$.  In an analogous way, we find $\Delta z = 0.117$ for $\log \NHI
\geq 12.5$ (which requires a smaller, higher S/N subset of the data). 
The absorption distance  $\Delta X=0.124$ \citep{Tytler87}, using the
prescription of \citet{Lehner07a},  which will be used for comparing \HI\
column density distributions in the next section.

\begin{figure}
%\epsscale{1.0}
%\includegraphics[width=185mm]{plot_nb_3c273_20091106.eps}
\includegraphics[width=85mm]{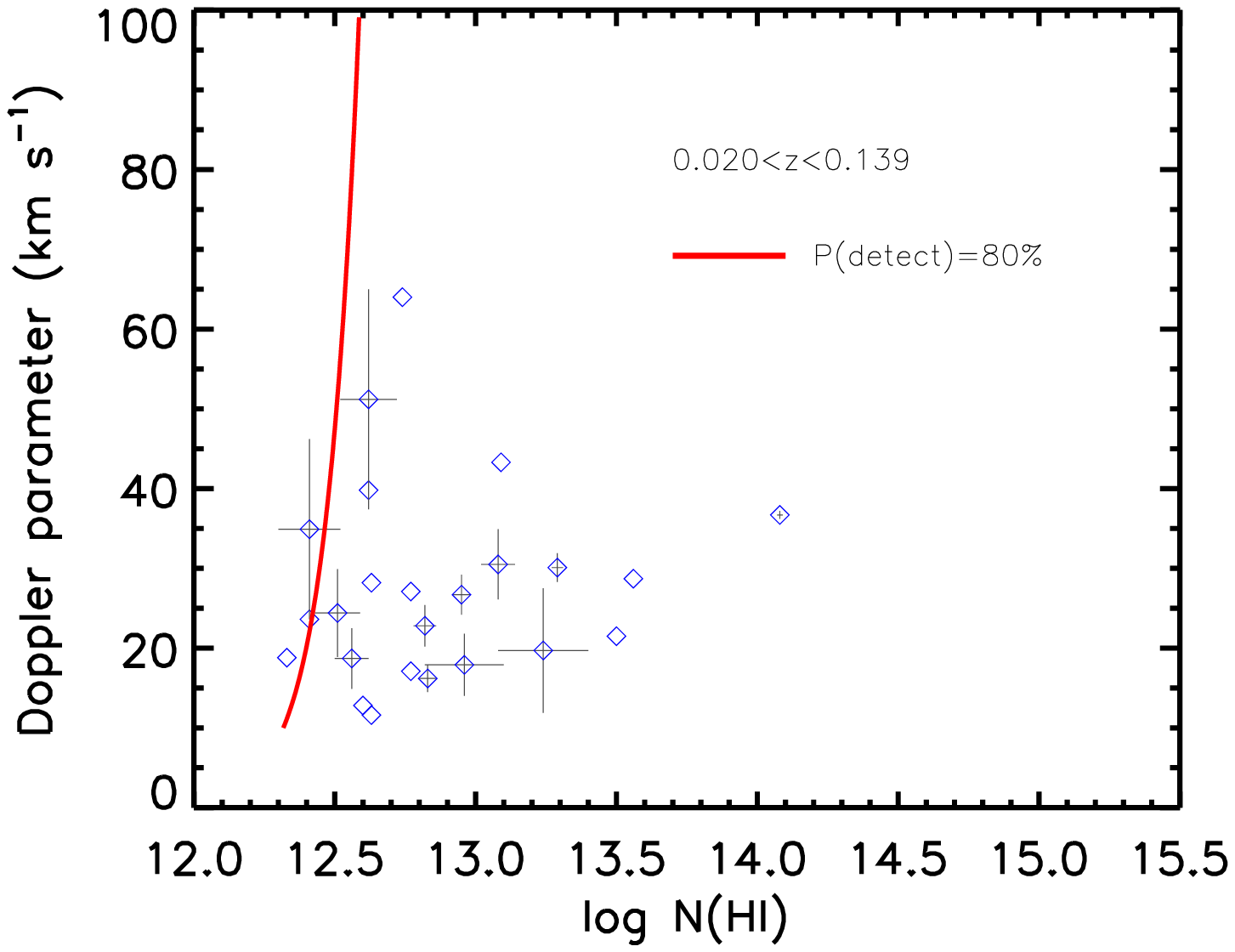}
\caption{\HI\ column densities and Doppler parameters. \Lya\ forest sample
over $0.020<z<0.139$, which is the redshift range used for our sample of
$\log \NHI\geq 12.5$.   {\it Open
diamonds:} data points. For clarity, only error bars for every second point
are shown.  The line indicates the parametric approximation to the 80\% detection
probability from  simulations in \S~\ref{sec:simulatedspectra}, based
on S/N$\geq 21$.  Three
absorbers with $12.3\leq \log \NHI \leq 12.5$ used for column density
studies are also included. \label{fig:nb}
}
\end{figure}

We list in Table~\ref{tab:linelist_lya} intervening
\HI\ and related metal redshifts, column densities and Doppler parameters,
including several systems either outside of the unblocked redshift 
intervals corresponding to the sensitivities mentioned above,  
or at $12.3\leq \log \NHI <12.5$ which nevertheless 
fulfill all of
the other selection criteria.
We show in Figure~\ref{fig:nb}
the Doppler parameters and \HI\ and column
densities for all \Lya\ forest systems that we detected 
using the procedure described in \S~\ref{sec:absorbersample}, overlaid with
the our parametrically fitted 80\% detection
sensitivities.

\paragraph{A. High column density lines $\log \NHI \geq 14.0$}

\begin{figure}
\includegraphics[width=180mm]{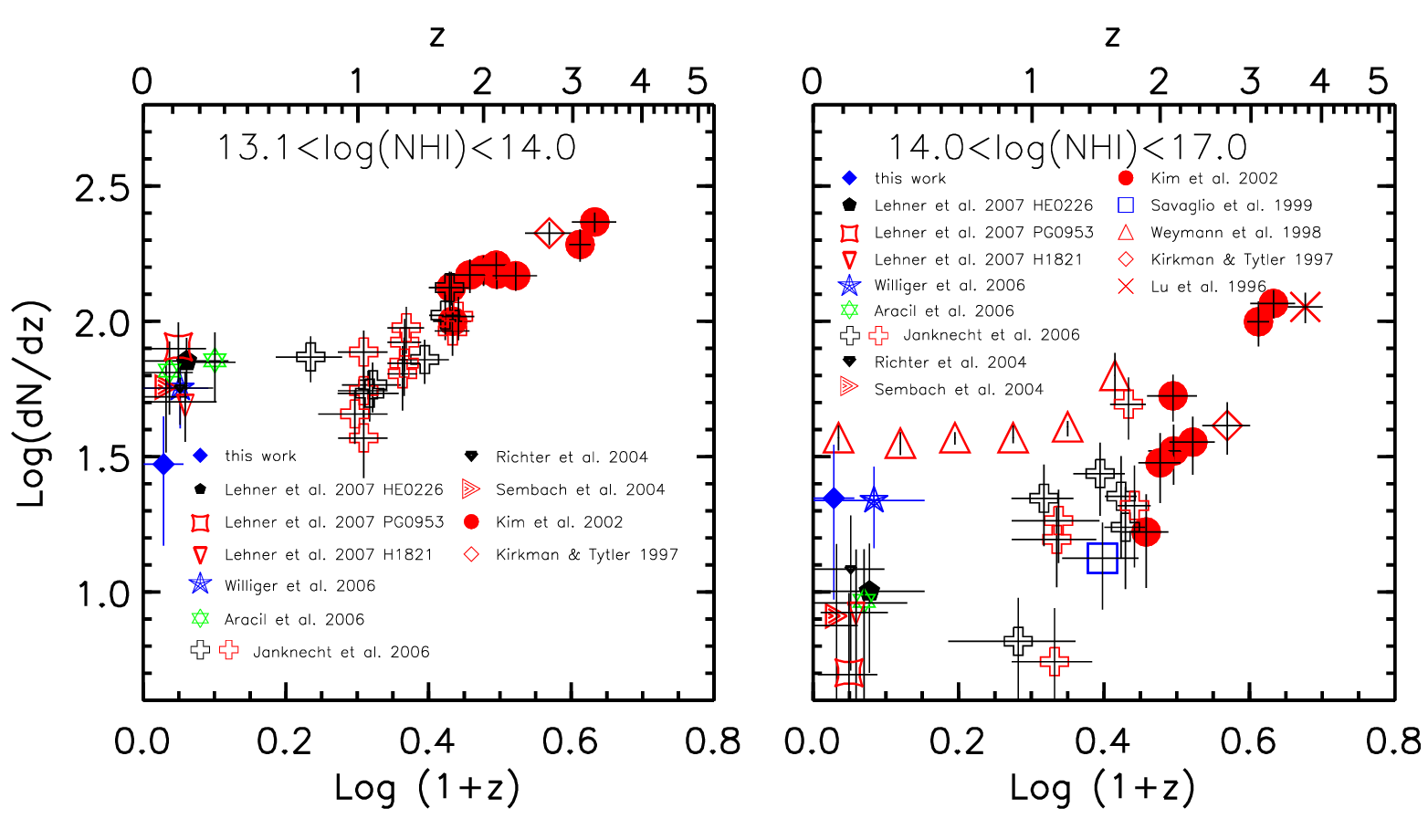}
\caption{{\it Left:} Ly$\alpha$ forest redshift density $dN/dz$ for
$13.1<\log  \NHI<14.0$. {\it Right:}  $14.0<\log \NHI<17.0$.   Error bars
show redshift bins and Poissonian errors in $dN/dz$. We restrict the data to
$b\leq 40$~\kms , and to errors in $b$ and $\NHI$ of $\leq 40$\% for 
3C~273 and the literature values in both cases except  for the
\citet{Weymann98} values, which are based on equivalent widths and
are shown for comparison. For the open crosses of \citet{Janknecht06}, dark symbols
denote the highest  quality data and light (red) symbols all data, as
defined in \S~4.2 of \citet{Lehner07a}. \label{fig:plotkim}
}
\end{figure}

For absorbers with $\log \NHI \geq 14.0$ from among our full sample of
21  systems, we find four over the interval $0.002<z<0.139$ ($\log dN/dz
= 1.47^{+0.18}_{- 0.30}$). If we restrict our sample to $b\leq 40$~\kms\
and errors in $b$ and \NHI\ to $\leq 40$\%, the  number reduces to 3
($\log dN/dz = 1.35^{+0.19}_{-0.38}$). The difference arises from  the
lower \NHI\ component of the $z=0.0053$ Virgo absorber, which has a
column  density error of a factor of 2 and is examined in  more detail in
\S~\ref{subsec:Virgoclustersystem}. We compare our results with the
literature (Fig.~\ref{fig:plotkim}),  and illustrate the scatter for
seven other sight  lines with contiguous redshift coverage of $\Delta
z>0.13$ from
\citet[][PKS~0405-123, $\Delta z = 0.413$]{Williger06},\footnote{We corrected
the line list for spurious absorbers, and will publish the updated list
in an erratum (Williger et al. 2009, in prep.).}
\citet[][PKS1116+215, $\Delta z = 0.133$]{Sembach04} and
\citet[][PG~1259+593, $\Delta z = 0.247$]{Richter04},
\citet[][HE0226-4110, H1821+643 and PG0953+415, $\Delta z =$ 0.397, 0.238, 0.202 
respectively]{Lehner07a},
\citet[][HS~0624+6907, $\Delta z = 0.329$]{Aracil06}.
The  redshift density is marginally high (similar
to PKS~0405-123) but still consistent with
other measurements from the literature shown in Fig.~\ref{fig:plotkim}
(right  panel),  similarly selected for $b\leq 40$~\kms and with errors
in $b$ and \NHI\ of $\leq 40$\%.   
The small number statistics and large cosmic scatter ($\sim
0.7$ dex in $dN/dz$ for individual sight lines) make make further analysis difficult
without larger samples.
The cosmic scatter range appears to
persist to $z\sim 1$. To illustrate the effect of the restriction in $b$
and in the errors, we also  included the redshift density from the large,
lower resolution  sample of \citet{Weymann98}, which was selected on the
basis of rest equivalent width and cannot filter out $b>40$~\kms\ 
systems, and is $\sim 1\sigma$ higher than our result. We note that the
$z\approx 0$ $dN/dz$ determination of \citet{Wakker09} is consistent with
the \citeauthor{Weymann98} result, as both  studies relied on equivalent
widths rather than column densities with restrictions on  the Doppler
parameters. The redshift densities for restricted, high resolution
$z\simlt 0.3$  sight lines are in general a factor of $\sim 2-3$ lower
than the analogous $1<z<2$ sight lines.

\clearpage
\begin{table}
\caption{Intervening absorption lines toward 3C~273 from STIS data
\label{tab:linelist_lya}}
%\tablehead{ 
%\colhead{species} & \colhead{$z$} & \colhead{error\tablenotemark{a}}     &\colhead{$b$ (\kms )} & \colhead{error\tablenotemark{a}} 
%& \colhead{$\log N$} & \colhead{error\tablenotemark{a}} 
%& \colhead{lines (\AA )/remarks\tablenotemark{b,c}}} 

\begin{tabular}{lccrrrcl} \hline
%\tablewidth{0pc}
%\tablehead{ 

species & $z$ & error$^a$ & $b$ (\kms ) & error$^a$ & $\log N$ & error$^a$ & lines (\AA )$^{b,c}$ \\ \hline

\footnotesize
%\startdata

$\!\!\!\!$~H I   &  0.002630 & 0.000039 &  125.8 &   44.9 &  13.04 &   0.14 & Ly$\alpha$     \\
 H I   &  0.003369 & 0.000002 &   37.4 &    0.8 &  14.22 &   0.02 & Ly$\alpha,\beta^e$	 \\    % OVI Sembach et al. 2001
 H I   &  0.005251 & 0.000021 &   26.1 &    1.3 &  14.33 &   0.31 & Ly$\alpha,\beta^e$	 \\
 Si II  &  0.005266 & 0.000005 &    5.2 &    2.6 &  11.78 &   0.09 & 1190,1260	  \\
 H I   &  0.005295 & \ldots   &   15.0 &    0.0 &  15.68 &   0.04 & Ly$\alpha,\beta, \gamma, \delta, \epsilon, \zeta, \theta, \kappa^e$	 \\
 C II  &  0.005295 & \ldots   &    8.6 &    0.1 &  12.80 &   0.07 & 1036,1334	  \\
 Si III &  0.005295 & 0.000002 &    8.2 &    2.1 &  12.36 &   0.07 & 1206, sets $z$ for Si~III, H~I     \\
 H I   &  0.007165 & 0.000029 &   38.8 &   12.1 &  12.67 &   0.12 & Ly$\alpha$     \\
 H I   &  0.007588 & 0.000037 &   64.0 &   18.0 &  12.86 &   0.10 & Ly$\alpha$     \\ 
% H I   &  0.011373 & 0.000026 &   27.6 &   11.5 &  12.43 &   0.14 & Ly$\alpha$     \\ DROPPED
% H I   &  0.016773 & 0.000027 &   28.3 &   11.5 &  12.31 &   0.14 & Ly$\alpha$     \\ DROPPED
% H I   &  0.017310 & 0.000025 &   32.7 &   11.0 &  12.42 &   0.12 & Ly$\alpha$     \\ DROPPED 27 JULY 2009
% H I   &  0.017699 & 0.000060 &   56.8 &   28.2 &  12.38 &   0.17 & Ly$\alpha$     \\
 H I   &  0.026225 & 0.000006 &   26.7 &    2.5 &  12.95 &   0.03 & Ly$\alpha$     \\
 H I   &  0.029348 & 0.000028 &   17.1 &    6.4 &  12.77 &   0.28 & Ly$\alpha$     \\
 H I   &  0.029471 & 0.000008 &   19.7 &    7.8 &  13.24 &   0.16 & Ly$\alpha^{d,e}$     \\
 H I   &  0.029578 & 0.000016 &   11.6 &    5.9 &  12.63 &   0.30 & Ly$\alpha$     \\
% H I   &  0.030479 & 0.000028 &   46.9 &   12.9 &  12.65 &   0.10 & Ly$\alpha$     \\
 H I   &  0.032798 & 0.000006 &   22.8 &    2.6 &  12.82 &   0.04 & Ly$\alpha$     \\
 H I   &  0.039373 & 0.000012 &   28.2 &    5.2 &  12.63 &   0.06 & Ly$\alpha$     \\
% H I   &  0.043792 & 0.000005 &    2.7 &    2.7 &  12.03 &   0.11 & Ly$\alpha$     \\
 H I   &  0.046639 & 0.000013 &   24.4 &    5.5 &  12.51 &   0.08 & Ly$\alpha$     \\
 H I   &  0.049000 & 0.000002 &   28.7 &    0.7 &  13.56 &   0.01 & Ly$\alpha^{d,e}$     \\
 H I   &  0.049953 & 0.000004 &   16.2 &    1.7 &  12.83 &   0.03 & Ly$\alpha$     \\
% H I   &  0.052367 & 0.000034 &   47.9 &   14.5 &  12.50 &   0.11 & Ly$\alpha$     \\
 H I   &  0.054399 & 0.000017 &   23.6 &    6.6 &  12.41 &   0.10 & Ly$\alpha$     \\
% H I   &  0.055635 & 0.000196 &  181.1 &  121.9 &  12.77 &   0.22 & Ly$\alpha$     \\
% H I   &  0.061502 & 0.000059 &   59.1 &   26.3 &  12.43 &   0.15 & Ly$\alpha$     \\
% H I   &  0.061923 & 0.000028 &   28.3 &   12.2 &  12.31 &   0.17 & Ly$\alpha$     \\
% H I   &  0.062296 & 0.000067 &   58.2 &   30.7 &  12.36 &   0.18 & Ly$\alpha$     \\
 H I   &  0.063510 & 0.000018 &   17.9 &    3.9 &  12.96 &   0.14 & Ly$\alpha$     \\
% H I   &  0.063660 & 0.000020 &   14.8 &    8.4 &  12.40 &   0.20 & Ly$\alpha$     \\
 H I   &  0.064078 & 0.000025 &   39.8 &   10.9 &  12.62 &   0.09 & Ly$\alpha$     \\
% H I   &  0.064373 & 0.000028 &   21.3 &   12.4 &  12.18 &   0.20 & Ly$\alpha$     \\
% H I   &  0.065697 & 0.000087 &  150.7 &   39.4 &  12.91 &   0.09 & Ly$\alpha$     \\
 H I   &  0.066548 & 0.000001 &   36.7 &    0.5 &  14.08 &   0.01 & Ly$\alpha,\beta^e$	 \\
% H I   &  0.067394 & 0.000068 &   70.5 &   35.1 &  12.53 &   0.16 & Ly$\alpha$     \\
% H I   &  0.067959 & 0.000036 &   28.3 &   14.3 &  12.26 &   0.21 & Ly$\alpha$     \\
% H I   &  0.068309 & 0.000044 &   52.6 &   22.0 &  12.65 &   0.18 & Ly$\alpha$     \\
% H I   &  0.068819 & 0.000063 &   80.8 &   29.4 &  12.79 &   0.13 & Ly$\alpha$     \\
% H I   &  0.069483 & 0.000034 &   48.5 &   14.8 &  12.57 &   0.11 & Ly$\alpha$     \\
% H I   &  0.069917 & 0.000038 &   47.6 &   17.1 &  12.51 &   0.12 & Ly$\alpha$     \\
 H I   &  0.073738 & 0.000034 &   64.0 &   15.0 &  12.74 &   0.08 & Ly$\alpha$     \\
 H I   &  0.074286 & 0.000033 &   51.2 &   13.8 &  12.62 &   0.10 & Ly$\alpha$     \\
 H I   &  0.075377 & 0.000009 &   27.1 &    3.8 &  12.77 &   0.05 & Ly$\alpha$     \\
% H I   &  0.078459 & 0.000067 &   97.8 &   27.7 &  12.69 &   0.10 & Ly$\alpha$     \\
% H I   &  0.079317 & 0.000055 &   65.2 &   22.9 &  12.54 &   0.13 & Ly$\alpha$     \\
% H I   &  0.079996 & 0.000054 &   51.0 &   22.0 &  12.41 &   0.15 & Ly$\alpha$     \\
% H I   &  0.080519 & 0.000030 &   34.2 &   12.7 &  12.33 &   0.13 & Ly$\alpha$     \\ eliminated June 2009
% H I   &  0.082594 & 0.000045 &  101.3 &   18.9 &  12.90 &   0.07 & Ly$\alpha$     \\
% H I   &  0.083166 & 0.000027 &   23.2 &   11.9 &  12.18 &   0.19 & Ly$\alpha$     \\
% H I   &  0.083923 & 0.000057 &   75.8 &   23.8 &  12.61 &   0.11 & Ly$\alpha$     \\
% H I   &  0.085312 & 0.000047 &   64.1 &   19.1 &  12.56 &   0.11 & Ly$\alpha$     \\
 H I   &  0.085882 & 0.000028 &   34.9 &   11.3 &  12.41 &   0.11 & Ly$\alpha$     \\
 H I   &  0.087632 & 0.000008 &   43.3 &    3.1 &  13.09 &   0.03 & Ly$\alpha$     \\
 H I   &  0.089746 & 0.000012 &   30.5 &    4.4 &  13.08 &   0.06 & Ly$\alpha$     \\
 H I   &  0.089898 & 0.000009 &   12.8 &    4.3 &  12.60 &   0.15 & Ly$\alpha$     \\
 H I   &  0.090110 & 0.000004 &   30.1 &    1.8 &  13.29 &   0.02 & Ly$\alpha^{d,e}$     \\
% H I   &  0.103050 & 0.000042 &   63.9 &   17.9 &  12.65 &   0.10 & Ly$\alpha$     \\ % eliminated June 2009
 H I   &  0.109141 & 0.000015 &   18.8 &    6.2 &  12.33 &   0.11 & Ly$\alpha$     \\
 H I   &  0.109380 & 0.000009 &   18.7 &    3.8 &  12.56 &   0.06 & Ly$\alpha$     \\
% H I   &  0.110262 & 0.000013 &   11.2 &    5.8 &  12.03 &   0.14 & Ly$\alpha$     \\
% H I   &  0.110771 & 0.000015 &    8.4 &    6.7 &  11.83 &   0.19 & Ly$\alpha$     \\
% H I   &  0.112197 & 0.000036 &   60.0 &   13.6 &  12.69 &   0.08 & Ly$\alpha$     \\ % eliminated June 2009
% H I   &  0.113862 & 0.000030 &   29.7 &   11.9 &  12.26 &   0.14 & Ly$\alpha$     \\
 O VI  &  0.12003\phantom{0} & \ldots  & \ldots  & \ldots  & 13.45 & 0.10 & 1032,1038$^f$ \\
 H I   &  0.120041 & 0.000002 &   21.5 &    0.7 &  13.50 &   0.01 & Ly$\alpha^{d,e}$     \\  %OVI, CIII system Sembach et al 2001
% H I   &  0.122425 & 0.000009 &    9.1 &    4.1 &  12.17 &   0.11 & Ly$\alpha$     \\
% H I   &  0.138549 & 0.000082 &  106.2 &   31.4 &  12.58 &   0.11 & Ly$\alpha$     \\
% H I   &  0.139566 & 0.000039 &   68.3 &   15.2 &  12.67 &   0.08 & Ly$\alpha$     \\
 H I   &  0.141676 & 0.000024 &   32.6 &    9.6 &  12.36 &   0.10 & Ly$\alpha$     \\
 H I   &  0.146575 & 0.000005 &   39.0 &    1.7 &  14.08 &   0.03 & Ly$\alpha,\beta^e$	 \\
% H I   &  0.149017 & 0.000018 &   16.5 &    7.2 &  12.00 &   0.13 & Ly$\alpha$     \\
% H I   &  0.149578 & 0.000048 &   49.1 &   18.1 &  12.22 &   0.13 & Ly$\alpha$     \\
 H I   &  0.150204 & 0.000021 &   32.9 &    7.8 &  12.37 &   0.08 & Ly$\alpha$     \\
% H I   &  0.151279 & 0.000053 &   57.5 &   20.2 &  12.32 &   0.13 & Ly$\alpha$     \\
% H I   &  0.155989 & 0.000139 &  101.3 &   53.9 &  12.60 &   0.20 & Ly$\alpha$     \\
 H I   &  0.157787 & 0.000006 &   21.5 &    2.1 &  12.63 &   0.03 & Ly$\alpha$     \\
 O VI  &  0.15779\phantom{0} & \ldots  & \ldots  & \ldots  & 13.39 & 0.11 & 1032$^f$ \\ \hline

%\enddata
%\vspace{2mm}
\medskip
\end{tabular}
$^a$ Errors are 1~$\sigma$, and assume that the component structure is correct.\\
$^b$ \phantom{.}$\!$Lines used in profile fits.\\
$^c$ We find an unidentified feature at 1149.3~\AA\ (also in the FUSE data).\\
$^d$ Inclusion of \Lyb\ makes no significant difference to the fit.\\
$^e$ \Lya\ absorption confirmed by \Lyb .\\
$^f$ \citet{Tripp08}.\\
%\tablenotetext{f}{Column density from \citet{Prochaska04}.}
%\tablenotetext{g}{The \CIV\ fit is from archival STIS G230M data.}
\end{table}

\clearpage

\clearpage
\clearpage

%{\it Low column density lines:} 

\paragraph{B. Medium column density lines  $13.1\leq \log \NHI < 14.0$}

In both in our full sample and  our restricted sample for $b\leq
40$~\kms\ and  errors in $b$ and \NHI\ of $\leq 40$\%, we find 4
absorbers over the interval $0.002<z<0.139$
with $13.1\leq \log \NHI <  14.0$ ($\log dN/dz =
1.47^{+0.18}_{-0.30}$). We compare our results with the  literature,  and
illustrate the scatter for seven other sight lines with contiguous
redshift coverage of $\Delta z>0.13$ from
\citet[][PKS~0405-123, $\Delta z = 0.212$]{Williger06},
\citet[][PKS1116+215, $\Delta z = 0.133$]{Sembach04} and
\citet[][PG~1259+593, $\Delta z = 0.247$]{Richter04},
\citet[][HE0226-4110, H1821+643 and PG0953+415, $\Delta z =$ 0.294, 0.238, 0.202 
respectively]{Lehner07a},
\citet[][HS~0624+6907, $\Delta z = 0.329$]{Aracil06}.
We note that \citeauthor{Lehner07a} estimated completeness to $\log\NHI =
13.2$, so the line densities for the three objects from that study 
should be a lower limit.  For PKS~0405-123,  we use a revised line list
for PKS~0405-123 (Williger et al.,  in prep.) and the redshift interval
known to be complete to $\log \NHI = 13.1$ (Paper~I).  We consulted with
N. Lehner about the signal-to-noise  characteristics of the other sight
lines in their work (private communication, 2007). We have split  the
HS~0624+6907 sight line into high and low $z$ halves to keep the
redshift  intervals to  similar length compared to that of 3C~273.
Results are plotted in Fig.~\ref{fig:plotkim} (left panel).  

The cosmic
scatter covers about 0.2~dex in  $dN/dz$ at $z\simlt 0.3$, for all sight
lines except 3C~273, which is  $\sim 1.5\sigma$ below  the mean of the
other seven sight lines.    For comparison, the $z\approx 0$ redshift
density of \citeauthor{Wakker09} of $\log dN/dz \sim 2\pm 0.1$ (based on
their fig. 5b)  is  just consistent with the maximum of the individual
sight line values, again possibly showing the difference in redshift
density  using equivalent widths vs. column densities with restricted
Doppler parameters. At $z\sim 1$, a larger sample \citep{Janknecht06}
indicates a cosmic scatter range of 0.35~dex which spans both the 3C~273
value  and the other seven low $z$ sight lines, so the 3C~273 sight line
may simply represent  the lower end of the redshift density range. The
high signal and high resolution in the data here make it unlikely that
blending and noise effects are mainly responsible for the paucity of
\Lya\ forest lines toward 3C~273 in this column  density range.   The
Cosmic Origins Spectrograph should reveal more definitively whether 
$\log dN/dz\approx 1.5$ is representative of one of the sparsest parts of
the \Lya\  forest, or if even more poorly populated sight lines exist.

\paragraph{C. Low column density lines  $\log \NHI \geq 12.5$}
\label{subsubsubsec:redshiftdensitylowcoldens}

Given the very high signal-to-noise ratio in the 3C~273 data, we also
calculate  the redshift density for $\log \NHI \geq 12.5$.  We find 21
absorbers above this threshold over $0.020<z<0.139$, for $\log
dN/dz=2.25^{+0.09}_{-0.10}$.  To compare the  3C~273 sample to the much
larger observational data set of \citep{Lehner07a}, we  consider
separately those systems with errors in Doppler parameter and \NHI\ under
40\%  (based on the lower limit $1\sigma$ errors for \NHI ) and
$b<40$~\kms .  These criteria focus the sample on relatively cool IGM gas
by excluding broad \Lya\ systems \citep[BLAs,
e.g.][]{Richter06a,Richter06b}, which are interpreted as the  signatures
of physically distinct low-$z$ hot IGM gas. The restricted 3C~273 sample
is 18  systems ($\log dN/dz=2.19^{+0.10}_{-0.12}$), which consistent to
$1\sigma$ in redshift  density with the full sample.

There are few comparisons down to a threshold of 
$\log \NHI \geq 12.5$ for absorbers at low
$z$. Observationally, \citet{Danforth08} found $\log dN/dz\sim
1.5^{+0.2}_{-0.1}$ for $z<0.4$ (see their fig.~3), which is  $2-3\sigma$
below our value of $\log dN/dz=2.19^{+0.10}_{-0.12}$, and
\citet{Wakker09} determined $\log dN/dz\sim  2.4\pm 0.2$ for $z\approx
0$, which is  consistent with our value.  These results could be 
understood in that the  weakest lines in the sample dominate the number
counts, based on the  column density distribution, and should affect an
equivalent width based sample (such as \citeauthor{Wakker09}) less than a
sample with a higher column density floor of $\log \NHI = 13.1$. From numerical
models, \citet{Paschos09} calculated $\log dN/dz\approx 2.1$ for the same
\HI\ threshold  at $z=0.1$, which is consistent with our measurements to
$1-1.5\sigma$. Less  directly, at $z=0$, our \hi\ column density
threshold corresponds to overdensities of $\log(\rho_H/\bar{\rho}_H)\sim
0.4\pm 0.2$ \citep{Dave99,DaveTripp01} as  calculated from hydrodynamical
models in the cold dark matter scenario. At $z\sim 3$, the  same
overdensity level corresponds to $\log \NHI \sim 14.0-14.5$, a rough
estimate  for which can be read from the right panel of
Fig.~\ref{fig:plotkim}: $dN/dz\approx  2.2\pm \sim  0.1$.  Our results
are therefore consistent with overdensities of
$\log(\rho_H/\bar{\rho}_H)\sim 0.4\pm 0.2$ within $1.1~\sigma$.  We
caution that  the correspondence of overdensity and $dN/dz$ may only
result from the considered   density perturbations being weak, as
non-linear growth would be expected over time to  skew the overdensity
distribution to high values; the low~$z$ IGM is relatively complex,
containing a significant fraction of shock-heated gas (which is related
to the  BLA population).

The observation of many more sight lines with COS would likely reveal 
better  the extent of cosmic scatter, and provide further comparisons
between the \Lya\  forest and galaxy environments e.g.  \citet{Chen05}.
In particular, COS could provide a  large increase in sight lines probed
to $\log \NHI \sim 12.5$, though at lower  resolution than this study.

\subsubsection{Doppler parameter distribution}
\label{subsubsec:dopplerparams}

\begin{figure}
%\plotone{q3c273_bhists_gmw.eps}
%\plotone{f3.eps}
%\includegraphics[width=185mm]{q3c273_bhists_gmw_20090801.eps}
\includegraphics[width=85mm]{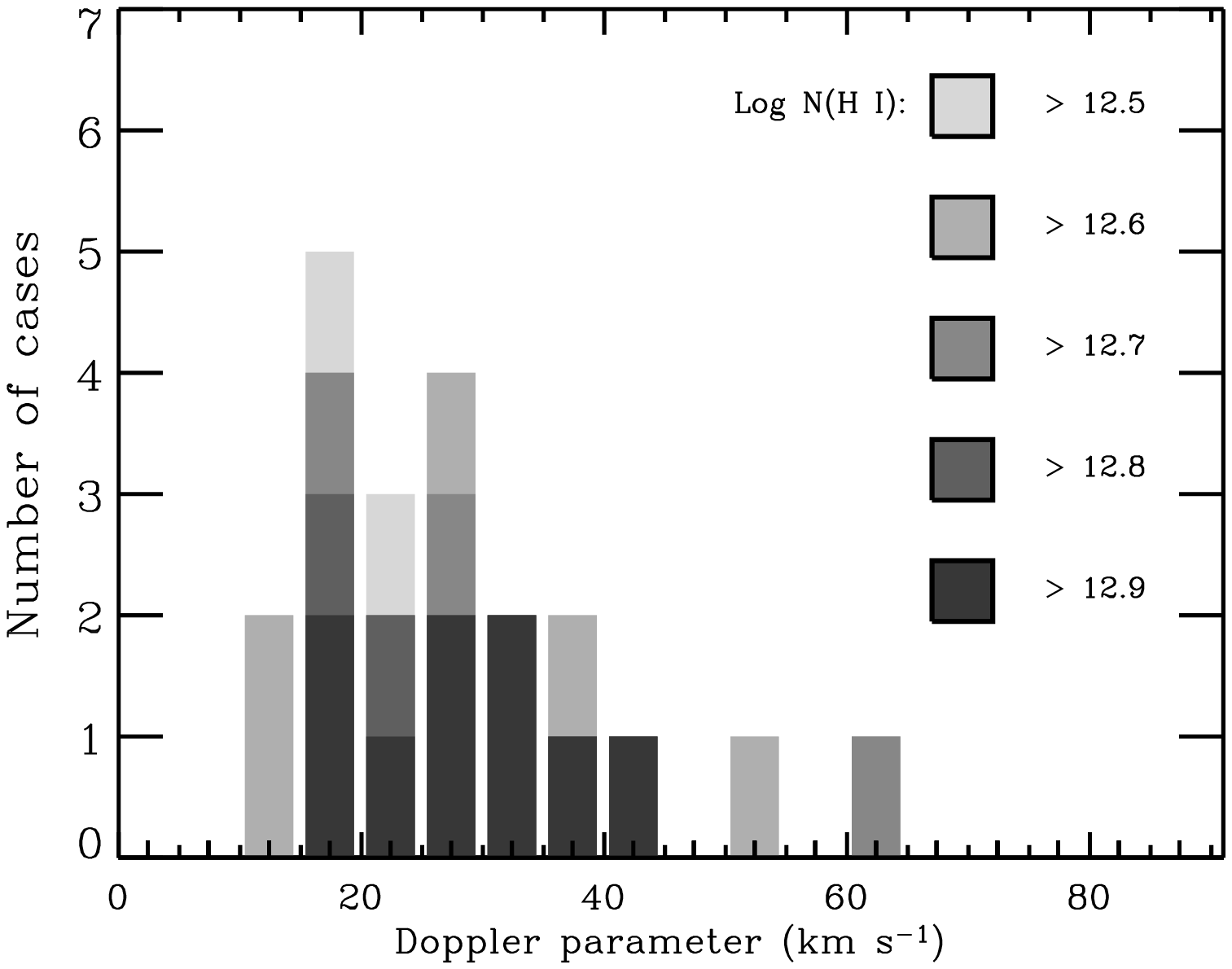}
\caption{Ly$\alpha$ Doppler parameter distribution for a series of minimum 
$\log \NHI$ thresholds in the complete sample.  \label{fig:dndb}
}
\end{figure}

The Doppler parameter distribution for the entire sample of 21 absorbers
has  mean, median and standard deviation of 28, 27, 13~\kms ,
respectively (Fig.~\ref{fig:dndb}).   In the same manner as
Paper~I, we tested whether  weak line blending is the cause. We divided
the strong sample into high and low \HI\  column density halves, which
occurs at  $\log \NHI= 12.80$,   and performed a Kolmogorov-Smirnov test
to determine the likelihood that the  Doppler parameters  are drawn from
the same distribution.  The probability that the samples are drawn  from
the same distribution is $P=0.96$.   The closest observed  \Lya\ forest
absorber  pair in velocity space is $\Delta v = 31$~\kms\ (between
$z=0.029471,0.029578$), or 4-5  times the resolution of the STIS data. 
The next closest pair is $\Delta v = 41$~\kms  .   This can be compared
to the mean velocity interval expected between absorbers  (in the absence
of clustering, which is expected to be weak and shows no evidence to the
contrary, \S~\ref{subsubsec:clusteringandvoids}) of $\langle \Delta v
\rangle  \sim 1500-1700$~\kms , based on our value of $dN/dz$ for $\log
\NHI \geq 12.5$ over $0.020<z<0.139$.  We would then expect to first
approximation that there is a $\sim$0.5\% chance of an absorber randomly being
$\sim 8-9$~\kms\ from another.  At such a velocity difference,
the lines could potentially be unresolved, depending
on the local S/N and the line column densities and Doppler parameters, with 
the greatest possibility of blending for broad, shallow lines  in relatively
noisy parts of the data. For the high S/N in the 3C~273 data and the
$b<40$~\kms\ absorbers with $<40$\% errors upon which we are focused, we
infer  that line blending is not significant for the \Lya\ forest  toward
3C~273.

We  compared results with the much larger low $z$ sample of
\citet{Lehner07a}.    They restricted their sample to $b<40$~\kms , to
correct for the non-Gaussianity of  the Doppler parameter distribution.  
If we make the same restriction for our data,  the sample would have 18
absorbers, with mean, median and standard deviation 24, 24  and 8~\kms .
\citet{Lehner07a} found values of 27, 27 and 8~\kms , which is  consistent
with our results.   The $\log \simgt 13.7$, $z<0.23$  sample of
\citet{Shull00} is also consistent (mean, median, standard deviation 31, 28, 7~\kms\
respectively). Finally, our $b<40$~\kms\ sample  is in accord with the
Doppler parameter mean of \citet{DaveTripp01} (25~\kms ), which is 
complete for $\log \NHI \geq 13.0$ and includes absorbers down to $\log
\NHI =12.6$, and thus shows agreement with  their
comparison model of  a $\Lambda$-dominated CDM universe \citep{Dave99},
and again with the more  recent simulations of \citet{Paschos09}.

%\clearpage
%\clearpage

\subsubsection{Column density distribution}
\label{subsubsec:columndensitydistribution}

\begin{figure}
%\epsscale{1.0}
%\plotone{plot_lehner_rev_3c273_08may.ps}
\includegraphics[width=180mm]{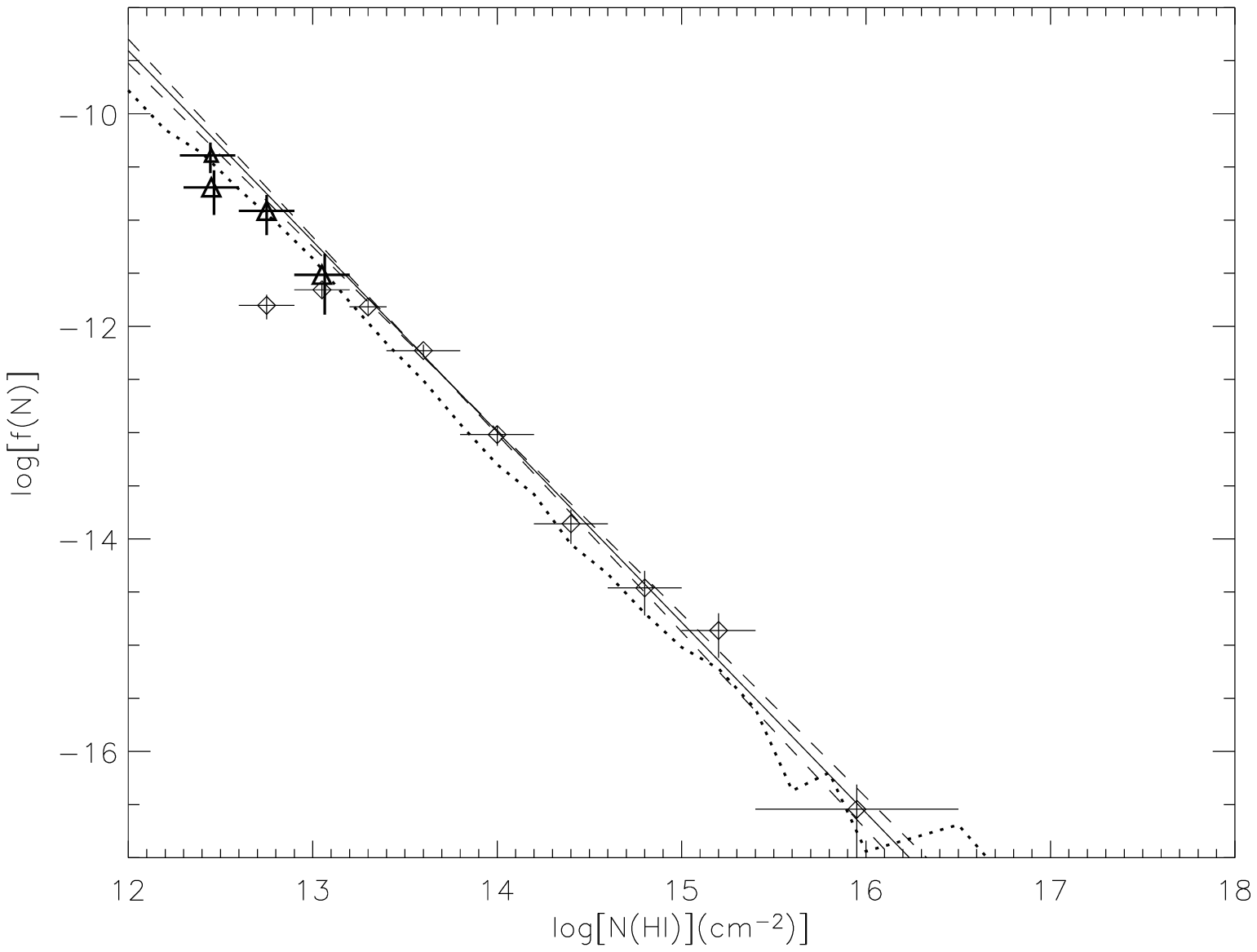}
\caption{\HI\ column density distribution, $\log[f(N)]=\log d^2N/dXd\NHI$.   We 
plot the  $\log \NHI \geq 13.2$, $b<40$~\kms , $\leq 40$\% error in $b$ and
\NHI\  subsample from \citet{Lehner07a}, recalculating the normalisation to
correct an error in  their fig.~14  \citep{Lehner07b}. 
{\it Diamonds:}  $dN/d\NHI$  for \citet{Lehner07a}\,  Their data are stated
complete to $\log \NHI \geq 13.2$,  with a turnover at lower \NHI .  {\it Large
Triangles:} 3C~273 data, with the same Doppler  parameter and \hi\ column
density restrictions (11 absorbers). The vertical bin centre bars for  the
$12.3\leq\log \NHI \leq 12.6$ and  $12.8\leq\log \NHI \leq 13.2$ bins have  been
slightly offset to the right for clarity.  {\it Small Triangle:} corrected value
for $12.3\leq \log \NHI<12.6$ bin (six  absorbers),  assuming a 50\% probability
of detection, estimated from our simulations.   {\it Solid line:} our best fit
slope to the complete \citet{Lehner07a} data only, $\log[f(N)]=A\NHI^{-\beta}$,
$\log A=12.11^{+0.90}_{-0.91}$, $\beta=1.793\pm  0.066$.  {\it Dashed lines:}
$1\sigma$ range in slope and normalisation.  We find the column density
distribution at $12.3\leq \log \NHI<13.2$ consistent at the  $\sim 1-1.5\sigma$
level with an extrapolation from that at $\log \NHI >13.2$.   A normalisation
error of 0.03 dex is negligible and not included.  {\it Dotted line:} Column
density  distribution from models of \citet{Paschos09}, table 4, for $z=0.1$.  A
normalisation  correction of $\log  f(N)+0.2$ (60\% increase) would best match
the data.
\label{fig:dndnhi}
}
\end{figure}

Observations suggest \citep[e.g.][]{Kim02,Misawa07} and simulations
predict \citep[e.g.][]{Paschos09}   that the column density distribution
should steepen at lower redshift for $\log \NHI  \simgt 13.0-13.5$, which
would be consistent with a systematic decrease of column  densities over
time for overdensities of a given strength. Our data can help to
investigate this trend, by uniquely probing the very lowest \HI\ column
densities in the \Lya\ forest at low $z$.  However, given the small
sample size and limited number of absorbers  with $\log \NHI \simgt
13.5$, we cannot obtain a good $\log dN/d\NHI$ slope  constraint using
just our data, in the framework of the commonly adopted parametric form
$dN/d\NHI \propto \NHI^{-\beta}$.   Therefore, we employ additional data
from  the literature to provide a larger comparison sample (larger $\Delta z$),
with the majority of absorbers at higher \NHI\ to extend the column density range. 
We used the  \Lya\ forest sample of \citep{Lehner07a}, which contains 270
\Lya\ systems at $z<0.44$, covering absorption distance $\Delta z =
2.40$, and a stated completeness to  $\log \NHI = 13.2$.  They calculated
the differential column density distribution for  a subsample 
with $b<40$~\kms ,  and errors in $b$ and \NHI\ to $<40$\%, for
a  total of 132 absorbers.  If we select from our sample in a way that is
consistent with these  rules,  the number of included 3C~273 absorbers at
$12.5 \le \log \NHI  < 13.2$ is  reduced to a subsample of 11.   We
calculate a maximum likelihood fit of $\beta=1.79\pm0.07$  for the
\citeauthor{Lehner07a} subsample over $13.2\leq \log \NHI \leq 16.5$, with
the distribution consistent with a power law to $\sim 2\sigma$
(Kolmogorov-Smirnov  probability $P=0.066$). This is consistent with the
\citeauthor{Lehner07a} value  of $\beta=1.76\pm0.06$. We wish to know
whether the weakest absorbers below the  completeness limit of
\citeauthor{Lehner07a} are consistent with a single power  law
distribution. The 3C~273 sample would comprise only 7.6\% of a combined
sample  with the \citeauthor{Lehner07a} data, over only 4.9\% of the absorption
distance, so it  would not contribute much weight to a combined sample.
We therefore simply extrapolate our best single power law fit  for the
\citeauthor{Lehner07a} subsample, and plot two data points from the
complete  3C~273 sample covering $12.5\leq \log \NHI \leq 13.2$ in
Fig.~\ref{fig:dndnhi}. The  largest  deviation shown by the 3C~273 data
is in the $12.5\leq \log \NHI<12.8$ bin,  which is within 0.8$\sigma$ of
the extrapolated fit.  The low column density end of the  column density
distribution is thus consistent with that at $13.2\leq \log \NHI < 
16.5$.

We can probe to 0.2 dex lower column density by correcting for the probability 
of detection  at $12.3\leq \log\NHI<12.5$ of $\sim 50$\%.  Using the same
selection criteria as \cite{Lehner07a}, there are 3 absorbers at $0.020<z<0.139$
toward  3C~273 in this extended column density range.  We plot both the observed
point, which  shows signs of a turnover in $dN/d\NHI$, and a factor of two
correction, which is  still consistent to $\sim 1.5\sigma$  at $12.3\leq
\log\NHI<12.5$ within the  extrapolated slope from \citeauthor{Lehner07a}.

\citet{Penton04}  found $\beta=1.65\pm 0.07$ (consistent to $1.5\sigma$ with our
results) over $12.3\leq \log \NHI \leq 14.5$ at $0.002<z<0.069$,  based on rest
equivalent widths and an assumed Doppler parameter of $b=25$~\kms .
\citet{DaveTripp01} used simulations to correct for incompleteness to show no
break in the \HI\ column  density distribution to $\log \NHI = 12.6$, and
calculated a consistent slope to our  value.

\clearpage
\clearpage

\subsection{Broad lines}
\label{subsec:broadlines}

\begin{figure}
\includegraphics[width=85mm]{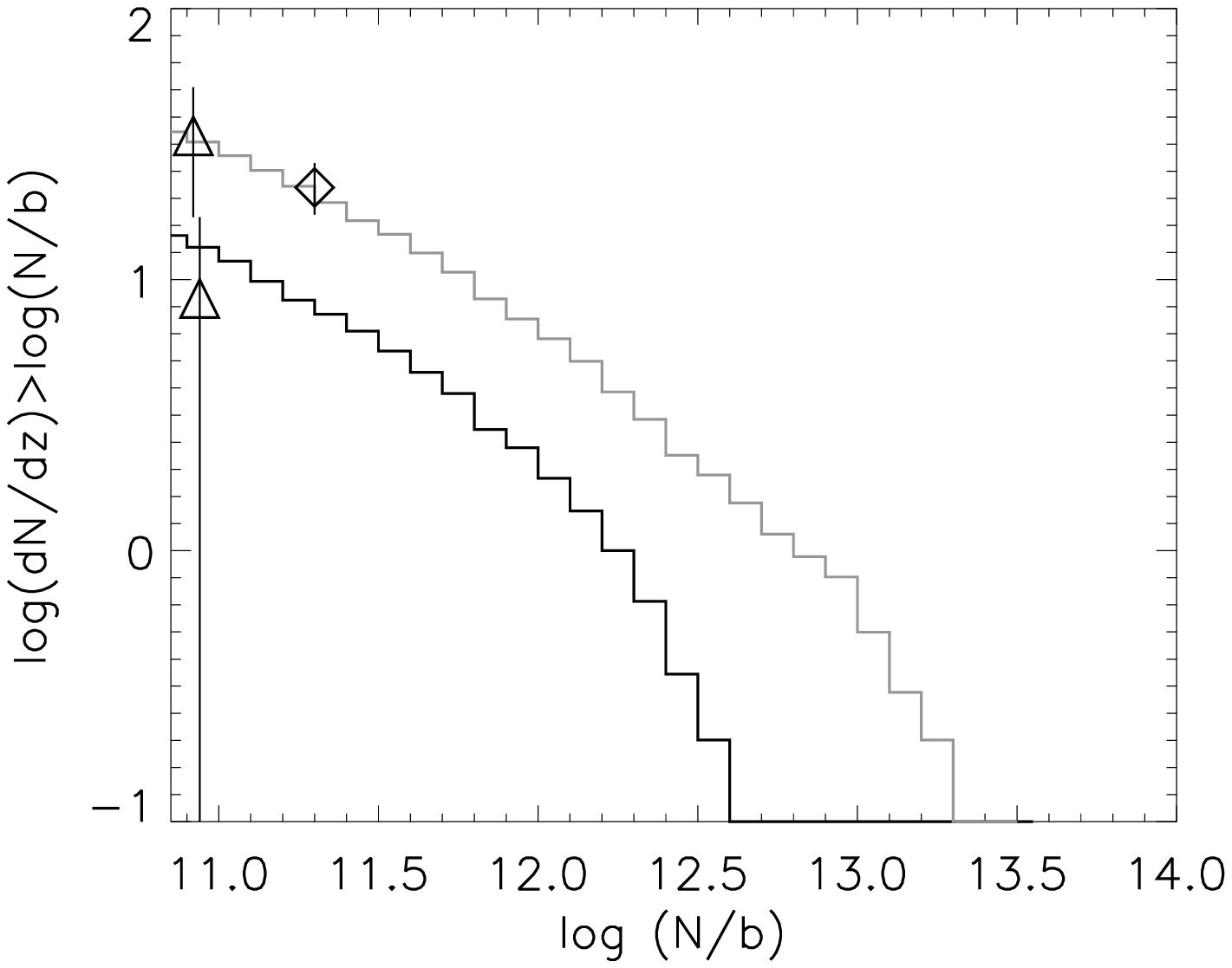}
\caption{Cumulative number of BLAs per unit redshift, 
$\log (dN/dz)_{\mathrm  {BLA}}$, as a function of the absorption strength,  
$\log (N({\mathrm {cm}}^{-2})/b({\mathrm {km\, s}}^{-1}))$ 
modelled after \citet{Richter06a}. The total
sample is shown in grey and the high quality sample is plotted in  black.
The diamond shows the value for $\log (dN/dz)_{\mathrm {BLA}}$ from 
\citeauthor{Richter06a} The triangles show the data from this study, both
the total sample (upper) and  the secure sample (lower).  Vertical error
bars have been slightly offset for  clarity.
\label{fig:plot_bla}
}
\end{figure}

Broad \Lya\ absorbers (BLAs), with $b>40$~\kms ,  have been a subject of 
interest in the past few years \citep[e.g.][and references
therein]{Bowen02,Penton04,Richter04,Sembach04,Richter06a,Richter06b,Wakker09}, 
because they may be the sign of a significant reservoir  of warm-hot IGM (WHIM)
gas.   Following the convention of \citeauthor{Richter06b}, we use a sensitivity
limit  of  \begin{equation} \frac{\NHI}{b}\simgt  \frac{3\times
10^{12}}{\mathrm  {(S/N)}}{\mathrm {cm}}^{-2}{\mathrm {(km\, s}}^{-1})^{-1} 
\approx 7.9\times 10^{10}  \end{equation} using the signal to noise ratio per
(their) 10~\kms\ resolution element in the above expression. This yields $\log
\NHI/b \approx 10.9$ given our minimum signal to  noise ratio per pixel of $\sim
21$ in the \Lya\ forest, $\sim 2$ pixel resolution  elements  and 3.2~\kms\
pixel size. Our BLA sample contains three \Lya\ absorbers in total ($z=$ 
0.07374, 0.07429,  0.08763), using the continuum generated by {\sc autovp} plus
manual modifications around  emission lines, and the compound feature detection
algorithm.   Assigning reliability to  BLA detections can involve subjectivity,
due to the nature of echelle spectra and  variety of detection and continuum
fitting techniques. Blending, low S/N ratio,  kinematic flows, Hubble broadening
and continuum uncertainties could be responsible for up  to $\sim$50\% of BLAs. 
To keep some subjective standards consistent between  different studies,  we had
one co-author (KS), a member of several collaborations which  have studied BLAs,
examine the BLA candidates.  He determined the $z=0.087632$  absorber to be
consistent with some ``reliable" samples of recent studies of BLAs e.g.
\citet{Lehner07a}.   Using this identification to divide the BLA sample into 
``total" and ``reliable", we find $dN/dz=25\pm 15$ for the total BLA sample and 
$dN/dz=8.5\pm 8.5$ for the reliable BLA, using a pathlength of $\Delta z=0.117$
(Fig.~\ref{fig:plot_bla}).   These are within $1\sigma$ of the values for the 
total and high quality samples in fig.~2 of \cite{Richter06a} ($dN/dz\sim 35$
and 15  for total and reliable systems, respectively), which were based on
simulated spectra  drawn from hydrodynamical simulations.

%\clearpage
%\clearpage

Given that the 3C~273 sight line is covered by the Sloan Digital Sky Survey
spectroscopic galaxy catalogue \citep[DR6,][]{Adelman-McCarthy08},  we examined 
the galaxy environment of all BLAs and \OVI\ systems beyond Virgo \citep[Virgo 
systems being studied in][, for example]{Tripp02,Stocke06,Wakker09} and  not in
the  3C~273 host cluster. The one secure BLA and one \OVI\ system each have several listed
galaxies with spectroscopic redshifts within  1.5~comoving Mpc impact parameter
and  $\Delta v \leq 200$~\kms .  Their \HI\ redshifts of $z=0.087632$ and
$0.090220$ correspond  to a line of sight distance of 9.8~local frame Mpc.

The $z=0.087632$ BLA has 2 galaxies with spectroscopic redshifts from the SDSS  
within 1.3 local frame Mpc  perpendicular to the line of sight ($\sigma_p$
plane) and $|\Delta v|\leq 75$~\kms\ ($\pi$ axis) \citep{Adelman-McCarthy08}.   
The next  nearest galaxy in the plane of the sky, in increasing velocity space
difference, is  2MASX J12271948+0200408 ($\sigma_p=2.95$~local Mpc,
$z=0.088579$,   $|\Delta  v|=261$~\kms ). The next closest galaxies within
1.6~local Mpc on the sky are three at  $z=0.0902\pm0.0002$, which are close in
redshift to the \OVI\ system (see below). We chose 10,000 random points
at $0.085\leq z \leq 0.095$ over $1^\circ \leq \delta \leq  3^\circ$ in
declination and $145^\circ\leq \alpha \leq 215^\circ$  in right ascension, which
contains 1825 galaxies allowing a $1^\circ$ border.    The SDSS spectroscopic 
sample covers $15.0<r<17.8$ in Petrosian $r$ magnitude, which corresponds to
$0.4-5L^*$  over $0.085<z<0.095$
\citep[$M^*_r=-21.4,-21.6$][]{Blanton01,Nakamura03}. By using the SDSS  as its own control
sample, systematic undercounting of dense regions due the maintenance of  a
minimum fiber distance should affect both the observed and the control samples
similarly.  We found  that a randomly placed absorber would have at least 2 SDSS
galaxies within 1.3~Mpc and $|\Delta v|\leq 75$~\kms\ 
with a probability of $P=0.006-0.008$, using
the  above search regions and varying the redshift limits by $\pm 0.025$.  There
is a third galaxy with a redshift in the NASA Extragalactic Database but not in
the SDSS,  LDTA 139870, which is 2.2 local frame Mpc from the line of sight with
$\Delta  v=7$~\kms ;   if it were in the SDSS, the random probability of finding
three galaxies within 2.2~Mpc would be the same.

The $z=0.090220$ \OVI\ absorber  \citep[$\log \OVI = 13.18\pm0.06$,][]{Tripp08} 
has 3 SDSS galaxies within 1.6 local  frame Mpc perpendicular to the line of
sight and $|\Delta v|\leq 50$~\kms .     The next closest galaxies within
2.6~local Mpc on  the sky are the three at  $z=0.0876\pm0.0002$ discussed with
the BLA above, and  2dFGRS~N387Z108 ($\sigma_p=2.84$~local Mpc, $z=0.0920$,
$|\Delta v|=489$~\kms ).  We chose 10,000 random points as for the $z=0.087632$
BLA, and found \ that a  randomly placed absorber would have at least 3 SDSS
galaxies within the same region with  a probability of $P=0.0004-0.0015$, using
the above search regions and varying the redshift limits by $\pm 0.025$. 
Properties for galaxies near this \OVI\  absorber and the reliable BLA are
listed in Table~\ref{tab:blagalaxies}.

\clearpage
\begin{table}
%\tablewidth{0pc}
\caption{Galaxies near $z=0.087$ BLA and $z=0.090$ \OVI\ absorber
\label{tab:blagalaxies}}
%\tablehead{ 
%\colhead{name} & \colhead{position (J2000)} & \colhead{$z$} & \colhead{$\Delta v$ (\kms )} & \colhead{$\Delta \theta'$} & \colhead{D$_\perp$}  & \colhead{r\tablenotemark{b}} & \colhead{references\tablenotemark{c}}\\
%\colhead{}    & \colhead{}                  & \colhead{}   & \colhead{from BLA/OVI\tablenotemark{b} }            & \colhead{}                 & \colhead{local Mpc}  & \colhead{}  & \colhead{}    }
%\startdata  
\begin{tabular}{lccccccl} \hline
%\tabletypesize{\scriptsize}
name                      &      position (J2000)    &      $z$   & $\Delta v$ (\kms )     & $\Delta \theta'$ & D$_\perp$ & r$^b$ & references$^c$ \\ 
                          &                          &            & from BLA/\OVI$^a$      &                  & local Mpc &       &                \\ \hline
%\startdata  
SDSS J122906.84+021348.2  &     12:29:06.8 +02:13:48 &   0.087361  & -75\phantom{-}&10.7	   & 1.05	&  17.0 &  	    2\\
LEDA 139870		  &     12:28:20.8 +02:21:34 &	 0.087660  & \phantom{7}7&  21.7           & 2.14       &  18.0 &              1,3\\
SDSS J122828.60+021109.3  &     12:28:28.6 +02:11:09 &   0.087717  &  23	&   12.4	   & 1.22	&  17.6 &	       1,3\\
2MASX J12285184+0206033   &     12:28:51.9 +02:06:03 &   0.090046  & -18\phantom{-}&\phantom{1}4.7 & 0.47	&  17.1 &	       1,2\\
2MASX J12280760+0202524   &     12:28:07.6 +02:02:52 &   0.090282  &  47	&   14.8	   & 1.50	&  17.3 &	       4  \\
SDSS J122806.93+015959.5  &     12:28:06.9 +01:59:59 &   0.090333  &  61	&   15.3	   & 1.54	&  16.7 &	       \\
%\enddata
%\vspace{2mm}
\medskip
\end{tabular}
%\tablenotetext{a}{Velocity difference from OVI system is for highest column density HI component.}
%\tablenotetext{b}{Magnitude from SDSS DR6 \citep{Adelman-McCarthy08}.  $M^*_r\approx -21.5$ at $z\sim 0.09$ \citep{Blanton01,Nakamura03},
%or $r\approx 16.5$.}
%\tablenotetext{c}{references: (1) \citet{Morris93} (2) \citet{Bahcall70}, 
%  (3) \citet{Paturel00}, (4) \citet{Tzanavaris06}}
$^a$ Velocity difference from OVI system is for highest column density HI component.\\
$^b$ Magnitude from SDSS DR6 \citep{Adelman-McCarthy08}.  $M^*_r\approx -21.5$ at $z\sim 0.09$ \citep{Blanton01,Nakamura03}, 
     or $r\approx 16.5$.\\
$^c$ references: (1) \citet{Morris93} (2) \citet{Bahcall70}, (3) \citet{Paturel00}, (4) \citet{Tzanavaris06}.\\
\end{table}

\begin{figure}
\vspace*{5cm}
\includegraphics[width=85mm]{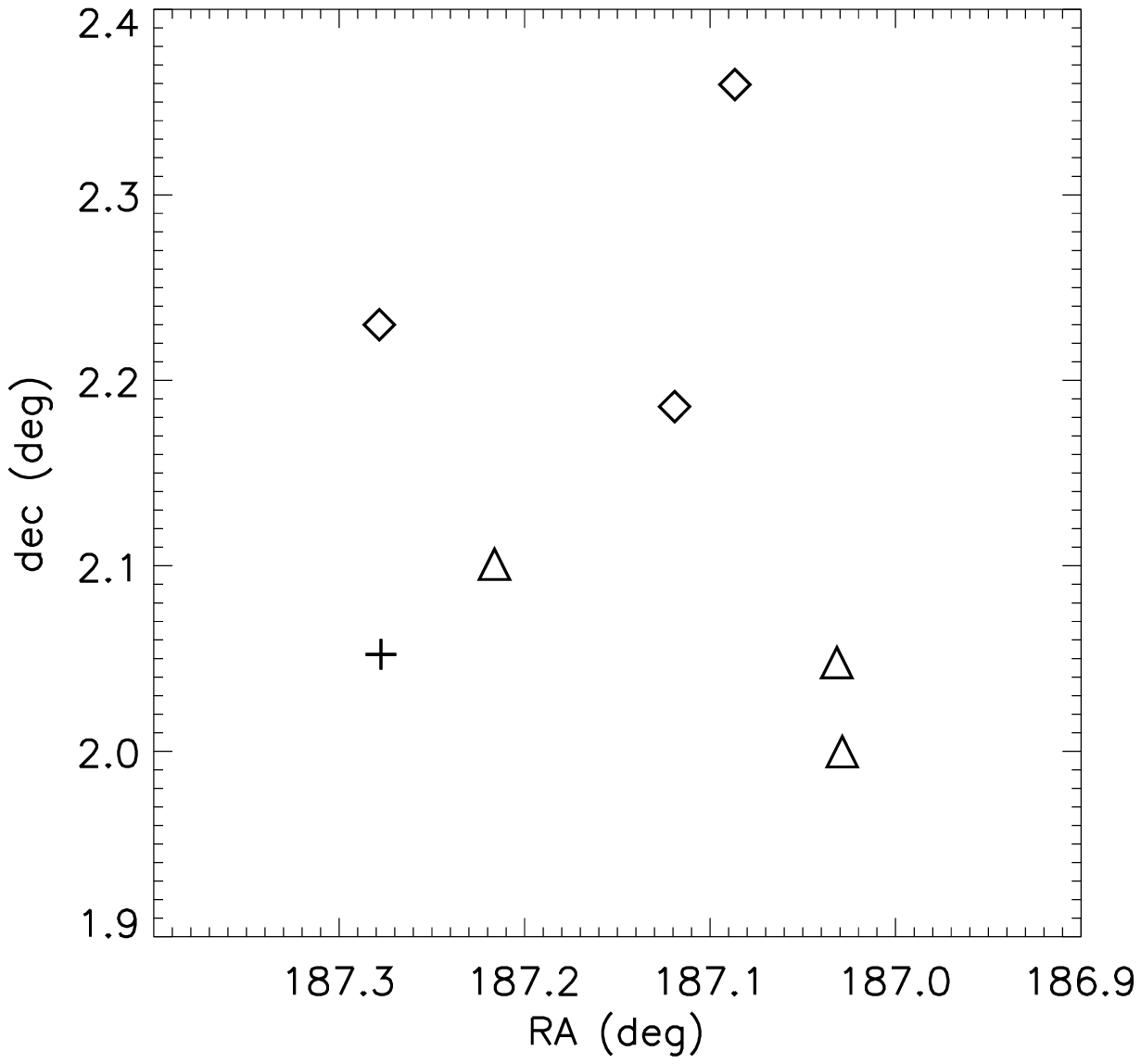}
\caption{Schematic of 3C~273 (cross) with galaxies within $|\Delta v|\leq 
200$~\kms\  of the $z=0.087632$ reliable BLA (diamonds) and $|\Delta v|\leq
50$~\kms\ of the $z=0.090220$ \ovi\ absorber (triangles). At $z=0.09$ in our
adopted cosmology, $0.1^\circ$ on the sky corresponds to 0.60 local-frame
Mpc.
\label{fig:plot_bla_galaxies}
}
\end{figure}

The closest three galaxies in velocity space to the $z=0.087$ BLA (including 
LEDA 139870, in the NASA Extragalactic Database but not in the SDSS
spectroscopic catalogue) form a rough rhombus with the 3C~273 line of sight,
implying that the  BLA gas is located in a sheetlike structure at least $\sim
1.1\times 2$~local Mpc in extent.  The multi-phase \OVI\ absorber  at
$z_{HI}=0.089898-0.09110$ is at  velocity separation $\Delta v=682$~\kms\ from
the $z=0.087$ BLA. The highest \HI\ column density component of the \OVI\
absorber is $-18<\Delta v<61$~\kms\ from the  three $z\sim 0.09$ galaxies, and
they trace an elongated formation $\sim 1.5$~local  frame Mpc long.   The six
galaxies are moderately luminous ($L/L^*\approx 0.5\pm  0.2$).   Based on SDSS
spectra and colours, and employing SED fitting from PEGASE \citep{Fioc97}, all
six galaxies near the BLA and \OVI\ absorber appear to be  moderate mass
($9.8<\log M/M_\odot < 10.9$), old ($\simgt 10$~Gyr) and with low star 
formation rates ($0.1-1.7M_\odot$~yr$^{-1}$). See
Fig.~\ref{fig:plot_bla_galaxies} for the relative positions of the 3C~273 sight
line and galaxies.  We also note that  there is a QSO/AGN close to that
redshift, 2MASx J12260241+0046403, which has $z=0.083$ ($\Delta v=1200$~\kms ), 
$b_J=18.57$ and is 89.3 arcmin or 8.4 local frame Mpc  away \citep{Hao05}, which
may be in a filament with the BLA. The fact that the two absorbers are separated
by $\sim 700$~\kms\ and that each is  on the scale of 1-2~Mpc from moderately
bright galaxies may  suggest a correlated physical environment for both
absorber-galaxy associations.

%\clearpage
%\clearpage

The BLA and \OVI -galaxy distances of 1-2~Mpc are consistent with the 
correlation lengths for such absorbers with $\sim 0.1-1.0L^*$ galaxies found
from  simulations by \citet{Ganguly08} but are larger than the 100-300~kpc
values found by \citet{Oppenheimer09}.   Observationally,  the velocity
differences between the galaxies and BLA+\OVI\ absorber are well within the
$\sim \pm300-500$~\kms\  differences used for much of the \citet{Wakker09} and
\citet{Stocke06} analyses, almost to the point that any putative absorber-galaxy
structures for our two  examples appear remarkably coherent over Mpc scales. The
galaxy-absorber distances  are a factor of several larger than the
few$\times$100~kpc distances found in a large, $z<0.15$ galaxy-\OVI\ system
correlation study by \citet{Stocke06}, and also are  large compared to the
350-450~kpc  range for \OVI\ absorbers with  $L\geq 0.1L^*$  galaxies found by
\citet{Wakker09} at $z\approx 0$.   \citeauthor{Wakker09}, \citet{Prochaska06}
and \citet{Cooksey08} found  significant variations in the galaxy environment of
\OVI\ absorbers.  The proximity of the secure BLA  and \OVI\ absorber with each
other and with luminous galaxies is consistent with the observation of 
analogous $z\sim 0.06-0.08$ \ovi -galaxy correlations in large- scale filaments
\citep{Tripp06} and the WHIM origin of BLA systems
\citep[e.g.][]{Richter06a,Richter06b}.

Finally, we compared the general correlation between local galaxy density and 
integrated \NHI\ column density for the 3C~273 sight line as in \citet{Bowen02},
and made a  Spearman rank correlation and Kendall $\tau$ correlation test
similar to the one in  Paper~I.  We used 43 galaxies from the NASA Extragalactic
Database within $2^\circ$ and 2  local frame Mpc  of the 3C~273 sight line at
$0.020<z<0.139$, where our \Lya\  detection probability is $\geq 80$\%. The
galaxy sample is inhomogeneous and  dominated by  SDSS entries, but with no more
significant selection effects  than the sample used in Paper~I.   We do not
probe further out in angular distance due to incomplete  coverage in the SDSS
DR6. We find no evidence of a correlation between galaxies and   column density
using absorbers with $\log \NHI \geq 12.5$.  The Spearman rank  coefficient for
redshift bin $\Delta z=0.003$ ($\Delta v\sim 800-900$~\kms )  is $\rho = 0.283$ 
with probability $p_\rho = 0.085$, and the corresponding Kendall rank
correlation is $\tau=0.250$ with $p_\tau=0.027$. Results are insensitive to
binning on larger  scales, and to lowering the redshift cutoff to $z=0.11$,
because there is only one  additional galaxy at $z=0.138$ below the end of the
\Lya\ forest sensitivity limit.   Smaller  scale bins are dominated by zero
counts of galaxies and \HI\ column densities.   This  lack of a correlation
contrasts with the weak correlation between galaxy number  density and summed
\NHI\ column density on 4000--6000~\kms\ scales detected toward PKS~0405-123 in
Paper~I.

The 3C~273 sight line should be probed to $\simlt 0.1L^*$ up to the QSO
redshift  to determine whether less luminous galaxies at smaller distances can
be associated  with the \OVI\ and BLA absorption. Larger cross-correlation
samples of BLAs+galaxies and BLAs+\OVI\ systems are needed to quantify further
any such  trends.

\subsection{Clustering and voids}
\label{subsubsec:clusteringandvoids}

%{\it Clustering:} 

\subsubsection{Clustering}
\label{subsubsec:clustering}

Clustering in the low $z$ Ly$\alpha$ forest has been studied in a number of 
cases \citep[e.g.][and references therein]{Janknecht02}, and  weak clustering
has been  indicated on velocity scales of $\Delta v \simlt 500$ \kms .   We
found a clustering  signal on the $\Delta v\leq 250$~\kms\ scale  with STIS
E140M data covering $z\simlt 0.4$ in a sample of 60 \Lya\ absorbers with $\log
\NHI \geq 13.3$ toward PKS~0405-123 (Paper~I).    The  clustering toward 3C~273
is expected to be weaker than for PKS~0405-123 due to  the  lower \NHI\
threshold. In a manner similar to Paper~I, we established a minimum line
separation of $\Delta v = 31$~\kms , based on the observed line separation
distribution in the actual data, and made 10000 random simulations of the 3C~273
line list for comparison to the data, assuming $dN/dz\propto (1+z)^\gamma$,
$\gamma=0.26$ \cite{Weymann98}.

\begin{figure}
\includegraphics[width=90mm]{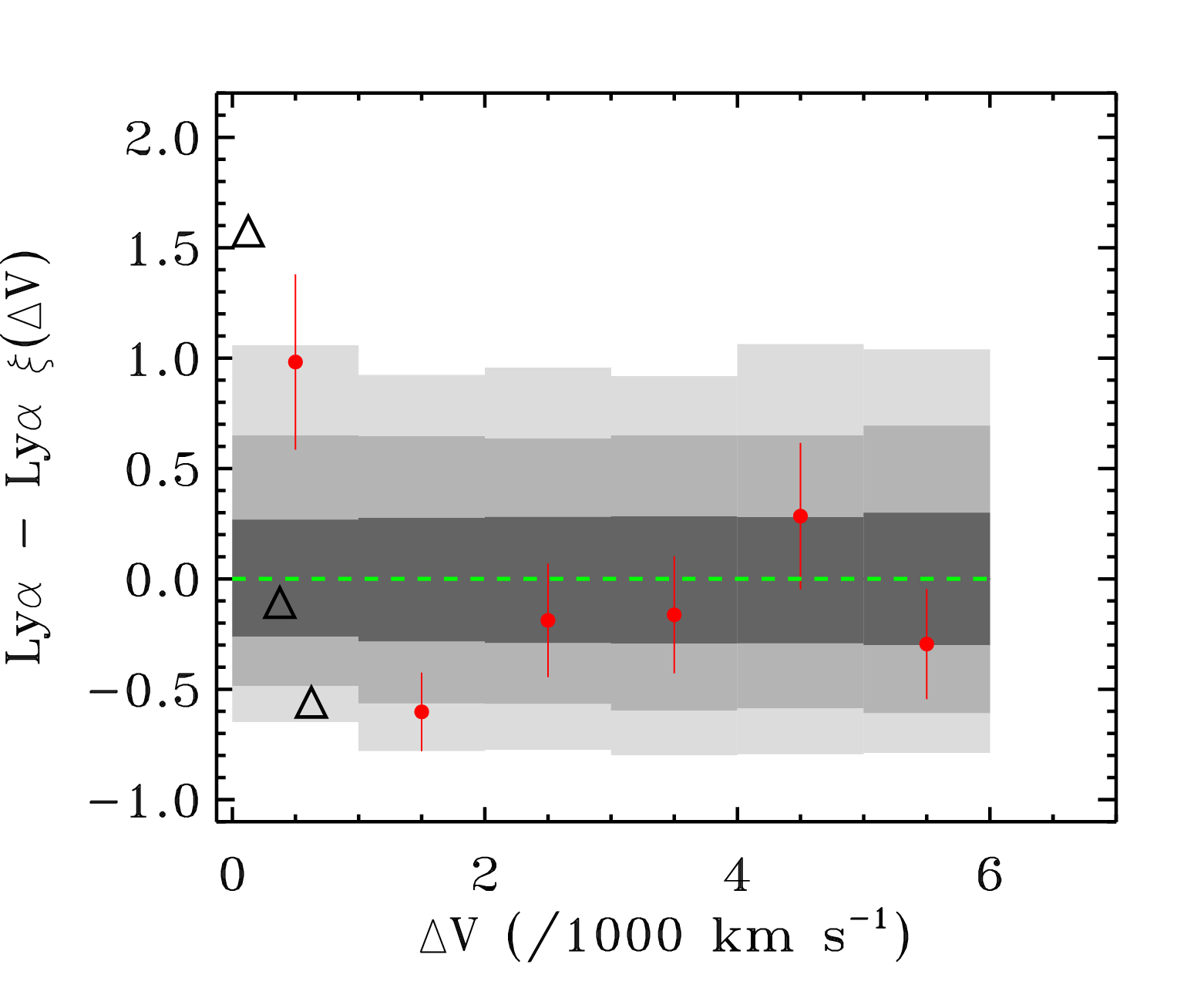}
\caption{Two point correlation function for the \Lya\ sample ($\log
\NHI\geq  12.6$) with $\Delta v=1000$~\kms\ binning, which shows the
strongest correlation in our sample, ($3.4~\sigma$). Shaded regions denote
68, 95, 99\% confidence limits from 10$^4$ Monte Carlo  simulations.
Simulated line lists have been filtered to eliminate pairs  with velocity
splittings of $\Delta v <  31$~\kms , as in the data. Error bars show
example $1\sigma$ Poissonian errors. The triangles show expected values from
the simulations of \citet{Dave03} for the same column density threshold; the
simulation box size artificially lowers the value of $\xi$ in the higher two
velocity bins.    
\label{fig:twoptcorr} 
}
\end{figure}

We find  clustering at the $3.0\sigma$ significance level in the two point
correlation function $\xi(\Delta v)$ for the 21 main sample  absorbers  with
$\log \NHI \geq 12.5$ at $0.020<z<0.139$, on the scale of $\Delta v\leq
1000$~\kms .   The signal increases in significance to $3.4~\sigma$,
corresponding to a strength of $\xi(\Delta v<1000)=1.0$ for $\log \NHI \geq
12.6$.   Poisson counts of the pairs in the bin, which are only an approximation
as the pairs between bins are correlated,  would indicate $1\sigma$ errors of
$\sim 40$\% .  The simulations give a 99\% confidence upper limit of $\xi(\Delta
v<1000)=1.1$. Based on the 10000 simulated line lists, the probability of the
signal being matched or exceeded is $P=2.8$\%.  The signal then diminishes in
strength at even higher column density thresholds due to the declining sample
size. Using the same hydrodynamical simulations for comparison as Paper~I
\citep{Dave03}, we find the measured clustering limits  consistent with the
predictions (Fig.~\ref{fig:twoptcorr}).   The $\tau$-step correlation function
\citep{Kim02} shows no significant signal.   The function appears sensitive
to rare high column density systems in an otherwise small sample dominated by
weak lines, as the only feature of note was contamination at the $\Delta v\sim
70-90$~\kms\  scale for $\tau\geq 0.4$ from the $z=0.066548$ absorber with $\log
\NHI=14.08$. The restricted \Lya\ sample with $b<40$ \kms\ and errors in $b$ and
\NHI\ of $\leq 40$\% only contains 16  systems, and is too small for useful
clustering analysis.

%\clearpage
%\clearpage

Our weak signal in the low $z$ $\log \NHI \simgt 12.5-12.6$  \Lya\ forest two
point correlation function  for $\Delta v < 1000$~\kms\ is consistent with the
weak clustering  predicted by numerical models.  However, it is on a
larger scale than the signal found toward PKS~0405-123 ($\Delta v<250$~\kms )
for $\log \NHI \geq 13.3$ On a similar $\Delta v=1000$~\kms\ scale,
\citet{Bowen02} found that \NHI\ column densities summed on 1000~\kms\ scales
correlate well with the volume density of $M_B\leq -17.5$  galaxies, which was
reminiscent of the effect we found  in Paper~I, albeit on 4000-6000~\kms\ scales.

There have been no comparable clustering studies in the low redshift \Lya\
forest to column density thresholds as low as
$\log \NHI = 12.5$.   Weak clustering has been found in the low $z$
\Lya\ forest for stronger absorbers on scales of up to 500~\kms.
\citet{Kirkman07} noted non-zero autocorrelations  in 74 FOS spectra with
230~\kms\ resolution at $z<1.6$ out to $\Delta v<500$~\kms\ using pixel 
opacities, which were strongly peaked at $\Delta v<100$~\kms\ for $z<0.5$ and
more broadly peaked at $0.5<z<1.5$. Their column density limits arise from sight
lines with a variety of sensitivities, but are presumedly $\simgt 0.5-1.0$ dex
higher in column density 
than here. \citet{Janknecht06} noted weak ($\xi(\Delta
v)\approx 0.1-0.4$),  marginal signals  on scales of $100<\Delta v<200$~\kms\
and $1000<\Delta v<2000$~\kms\ for $12.9<\log \NHI<14.0$. \citet{Penton04}
determined a  two point correlation function signal  of $\xi(\Delta v<190~{\rm
km\,s}^{-1})\sim 3.3$  at the $4.5\sigma$ significance level  and  $\xi(\Delta
v<260\,{\rm km\,s}^{-1})\sim 2.8$ at  $5.6\sigma$ significance for rest
equivalent width  $W_0\geq 65$~m\AA\  ($\log \NHI\simgt 13.1$  for $b = 25$~\kms
).   They also found a  $3\sigma$ excess at $260<\Delta v<680$~\kms .  However,
they found no significant correlation for weak absorbers ($W_0< 65$~m\AA ) with
either themselves or with stronger absorbers, which is consistent with our
results.    Our 99\% confidence limit of $\xi(\Delta v)=1.1$ at $\Delta
v<1000$~\kms\ is at least crudely  comparable if not a slightly lower constraint
than their four $1\sigma$ upper limits of $\xi(\Delta v)\simlt 2$ over a similar
velocity range. A sample size an order of magnitude larger (containing
100$\times$ more pairs and roughly reducing bin errors by a factor of 3) would
make either more interesting clustering  constraints for weak systems or detect
a correlation  strength predicted in our comparison models. COS data should
address this question.

\subsubsection{Voids}

As in Paper~I,  we searched for regions in velocity space empty of Ly$\alpha$
systems above a series of \hi\ column density thresholds, and compared them with
the frequency of similarly-sized gaps in randomised line lists having the same
statistical properties as the  observations.   The most significant gap  we find
has a line of sight extent of $\Delta v=8100$~\kms\ (120 comoving Mpc) at
$0.0901<z<0.1200$ for $\log \NHI \geq 12.6$.  The probability of such a gap
being matched or exceed in a sample with the same number of \Lya\ lines  is  
$P=0.006$, using simulated line lists having a Poisson mean of 21.  The void 
persists  at more marginal significance for $\log \NHI \geq 12.7$, with
$P=0.022$, and for higher \NHI\ thresholds the  sample becomes too small to be
statistically useful.  For $\log \NHI \geq 12.5$, the most significant void is
over $0.0072<z<0.0262$, with $P=0.023$.

We can use the SDSS DR6 spectroscopic sample \citep{Adelman-McCarthy08} as a
rough indicator of galaxy overdensity, although by $z\sim 0.12$ it is only
probing bright galaxies. We count 0,3,9,14 galaxies in the SDSS DR6 within
1,2,3,4 local-frame Mpc  respectively in the line of sight at
$0.0901<z<0.1200$.   We made counts of galaxies within a grid of 1-4 local frame
Mpc around 10000  random locations in the same redshift interval, over
$140^\circ < \alpha < 225^\circ$  and $- 0.25<\delta<3.0^\circ$, and found a
minimum random probability (for 0 galaxies out to a maximum radius of 1.48~Mpc)
of $P=0.88$ for the \Lya\ forest gap to be in a region up to similarly as
dense.  The maximum probability is $P=0.988- 0.989$ for radii $r=3,4$~local Mpc.
The galaxy environment therefore does not appear significantly underdense, and 
could in fact be overdense.  A comparison with the interval $0.08<z<0.09$, which
is comparatively rich in that it contains 3 absorbers and 0,2,4,6 galaxies 
within $r<1,2,3,4$~local frame Mpc, respectively, reveals no significant over or
underdensity. ($P$ ranges from 0.771-0.922.)  Deeper galaxy surveys may  reveal
more about the relationship between the weakest absorbers and nearby galaxies.
We do note from the NASA Extragalactic Database (NED) that a  QSO   $\sim
1^\circ$ from 3C~273, RX~J1230.8+0115 has V=14.4, rest-frame Lyman limit flux 
$F_{1018\AA }\approx 5\pm 1 \times 10^{-14}$~erg~cm$^2$~s$^{-1}$~\AA$^{-1}$ (\FUSE\
program 1099001, PI Sembach), $z=0.117$ and is 54.3 arcmin or 6.9 local frame
Mpc from the 3C~273 sight line  \citep{Read98}.  For comparison, the QSO 
SDSS~J030435.32-000251.0, which was suggested  as causing the $z=3.05$ \HeII\ opacity
gap toward Q0302-003 \citep{Jakobsen03,Heap00}, is  only 3.0 local frame Mpc from
the line of site and is likely brighter by several tenths of a  dex  than
RX~J1230.8+0115 ($u=22.35$; a full comparison is beyond the scope of this work).
The  $z=0.1200$ \OVI\ absorber  toward 3C~273 is only $\Delta v\sim 800$~\kms\
from  the RX~J1230.8+0115 redshift, and both objects may be located in the same 
filamentary structure.

Voids in the \Lya\ forest are rare, and most of the work in the field has been
done at high  redshift. Few \Lya\ forest void studies have been done at $z<1.6$,
but they are valuable as comparisons to  the galaxy-absorber relation with
respect to the inverse question of \Lya\ absorbers in galaxy voids
\citep[e.g.][]{Stocke07}. In Paper~I, we found toward PKS~0405-123  gaps at
$0.0320<z<0.0814$ of 206~comoving Mpc as defined  by $\log \NHI \geq 13.3$
absorbers with probability of random occurrence in shuffled data of $P=0.0004$,
at $0.1030<z<0.1310$ for $\log \NHI \geq 13.2$ (113~comoving Mpc, $P=0.007$)
and at $0.0320<z<0.0590$ for $\log \NHI \geq 13.1$ (also 113~comoving Mpc,
$P=0.003$). 
There are several higher redshift
voids ($z\sim 1.6-3.2$) on the $40-60$~Mpc scale for $\log \NHI \sim 13.4-13.5$,
which we discussed in Paper~I. Given the expected column density  {\it vs.}
density perturbation evolution between $z\sim 3$ and $z\sim 0.2$, the column
density levels for which we find our low redshift void(s) toward 3C~273 ($12.5
\leq \log \NHI \leq 12.7$) are comparable to  $\log \NHI \approx 14.0$ at $z=3$
(\S~ref{subsubsec:redshiftdensity}).   
Our 3C~273 data are thus $\sim 0.5$ dex too
shallow to probe similar density perturbations as used to define the gaps at
$z\sim 2-3$.  Therefore, we do not consider it unusual to find a slightly larger
candidate gap than reported at higher redshift, given the difference in column
density sensitivity. We do not find the $0.0901<z<0.1200$ region within radius 
$r\leq 4$~local frame Mpc of the 3C~273 sight line to be of unusually low galaxy
density,  based on the SDSS DR6 spectroscopic survey. A more definitive
measure of the galaxy environment would require a more sensitive survey.

Although the $0.0901<z<0.1200$ gap toward 3C~273 is coincident with
RX~J1230.8+0115 ($\Delta \theta=54.3$~arcmin or 6.9 local frame Mpc),  recent
analyses of any transverse proximity effect on smaller scales of $\sim 1-5$~Mpc
have not indicated significant thinning of the \Lya\ forest which could produce
voids.  Possible explanations for the lack of any detections include anistropic
emission \citep[e.g.][]{Hennawi07}, or the masking of any transverse proximity
effect by enhanced density environments around QSOs \citep{Goncalves08}.  The
correlation length between QSOs and optically thick \HI\ absorbers off the line
of sight is $9.2^{+1.5}_{-1.7}h^{-1}$~comoving Mpc at $z\sim 2.5$
(\citeauthor{Hennawi07}), which is three times stronger than the QSO-Lyman break
galaxy correlation, would be on the order of the observed distance between
RX~J1230.8+0115 and the $z=0.1200$ \OVI\ absorber  toward 3C~273.  Both objects
could therefore mark the same structure.  A deep galaxy survey toward the region
would confirm this.

\citet*{Shang07} searched the SDSS for gaps using a pixel opacity 
technique with limits of transmitted flux $F=0.6-0.8$ at $2.8<z<3.9$, which
probes rest equivalent widths corresponding to  $\log \NHI \approx 13.0-13.5$
for $18<b<45$~\kms , or overdensities of $\log \rho/\bar{\rho}=-0.2$.  They
found void sizes of $\sim 3-35$~comoving Mpc, which were in agreement with the
standard $\Lambda$CDM model for \Lya\ forest formation. As our \Lya\ absorber
sample does not probe down to the same overdensity level 
as the \citeauthor{Shang07}
search, our results are not inconsistent with theirs. Given the
probability of random occurrence of the largest gap we identified, the current
paucity of low redshift data sensitive to $\log \NHI = 12.5$ and the lack of
data which probe the same overdensities as \citeauthor{Shang07}, future studies
with COS should be used to confirm the frequency of large
voids.

\subsection{The Virgo cluster absorber at $z=0.00530$}
\label{subsec:Virgoclustersystem}

\cite{Tripp02} established metallicity limits and corresponding physical
conditions for the two Virgo cluster absorbers, including the one at
$z=0.00530$.  We took the opportunity to re-fit the components. We used nine Lyman
series lines through Ly$\kappa$ (919\AA ), and included data from  STIS and the
revised reductions from the
LiF1A, LiF2B, SiC2A and SiC1B \FUSE\ channel segments.  (The SiC segments
covering \Lyb\ were of such low signal that they did not help to give useful
constraints.)

\begin{figure}[h]
%\plotone{lya2pc.n125170.dv500.vsm27.06jan.mc1e4.ps}
%\epsscale{0.8}
\includegraphics[width=135mm]{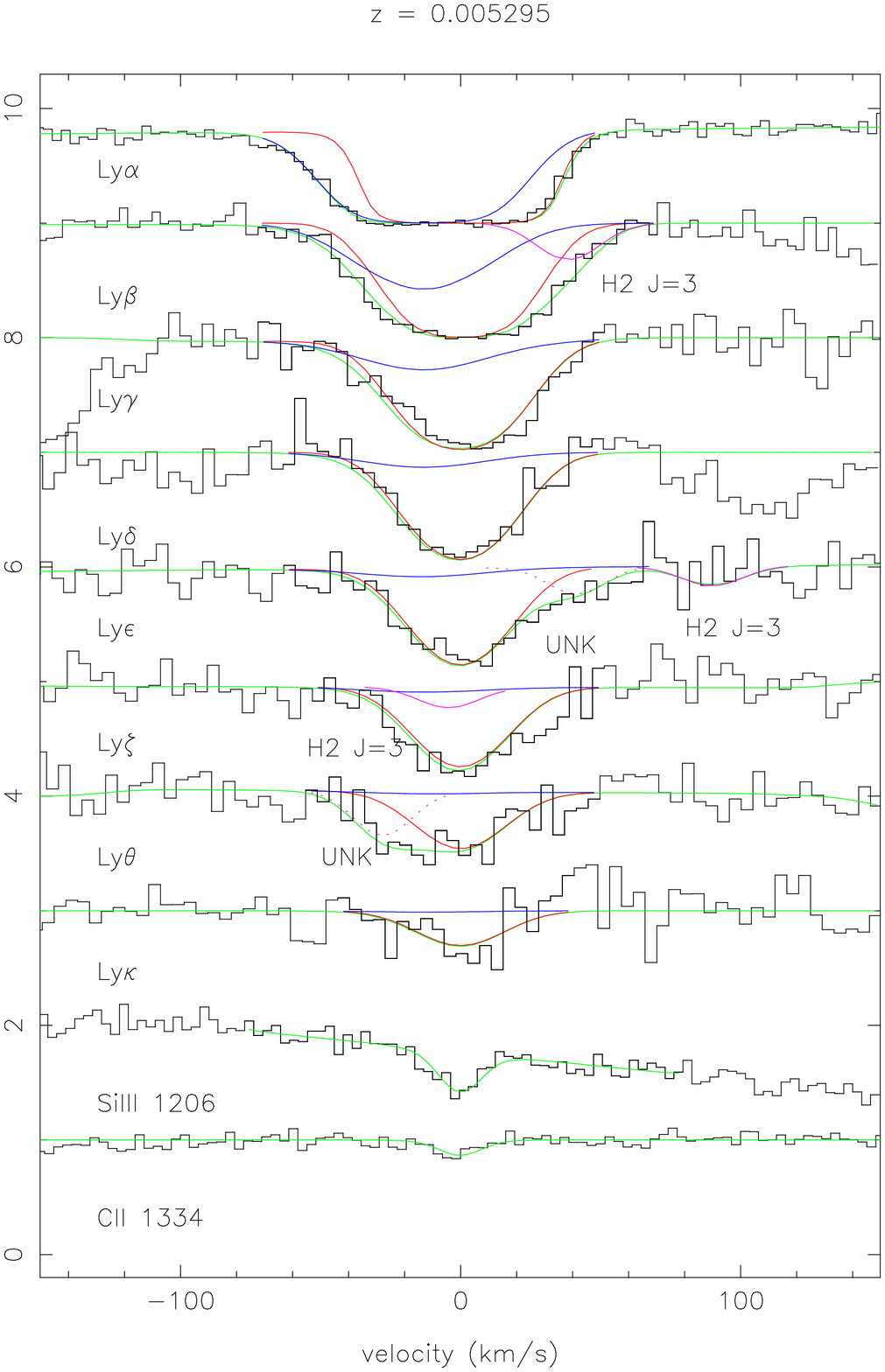}
\caption{Two component fit to $z=0.00530$ Virgo absorber, including
Si\,{\sc III}~1206 and \CII~1334.
Green - total profile.
Red   - higher \hi\ component, at the metal $z$.
Blue  - blue component, \hi\ only.
Lilac - blends: {\it solid} -- \Htwo ; {\it dashed} -- unknown (UNK).
Note that the colour sequence from top to underneath is lilac-blue-red-green
where there is more than one component (i.e. \hi , not metals) so sometimes 
green 
especially is obliterated by a component.
\label{fig:virgofit2}
}
\end{figure}

We took advantage of the clear \SiIII\ feature to set the redshift for the
main \HI\ component.  Two components 
provide a  satisfactory fit to the data (Fig.~\ref{fig:virgofit2}). The
reduced $\chi^2=1.19$, though the formal probability of a  fit to the data
is still $<1$\%. One component makes a significantly poorer fit, and three
are not substantially  better (reduced $\chi^2=1.234$). A four component fit
is marginally worse than the two component fit (reduced  $\chi^2=1.209$), 
but suffers from increasing uncertainties in the \HI\ column densities.  We
therefore adopt a two  component fit with a total \HI\ column density $\log
\NHI = 15.70\pm 0.04$ ($1\sigma$  error). This is within the $1\sigma$ error
of the one component fit from the curve of  growth measurement of $\log
\NHI=15.77^{+0.12}_{-0.10}$ excluding \Lya , which has the most profound
blending effect by the second component, by \citet{Sembach01}.
\citet{Tripp02} adopted $\log \NHI=15.85^{+0.10}_{-0.08}$, which was  the
\citeauthor{Sembach01} figure (including \Lya\ in the curve of growth fit)
for their metallicity  studies. The conclusions of \citeauthor{Tripp02}
about the nature of the $z=0.0053$ absorber should not be significantly
affected.  It is possible that the component structure is more complex, but
higher signal and resolution  far-UV data would be necessary to make a
significant improvement, which must wait for a successor to \FUSE.  

\clearpage
\clearpage

\section{Conclusions}
\label{sec:conclusions}

We have performed an analysis of the \Lya\ forest toward 3C~273 based on STIS
E140M and a re-processing and reduction of \FUSE\ spectra, and find the 
following.

1. We present STIS E140M echelle data for 3C~273, and performed profile
fits  or measured apparent optical depths for all of the detected \Lya\
forest over $0<z<0.158$,  plus  intervening metal and Galactic metal
absorption systems.  We analysed simulated spectra to determine our
sensitivity to line detections and resolution of close pairs, and probe to
$\log \NHI \approx 12.5$ over most of the spectral range. Our main sample
consists of 21 absorbers to $\log \NHI \approx 12.5$ over $0.020<z<0.139$.

2. The redshift density for the absorbers with column densities $\log \NHI>
14.0$  is consistent with previous, lower resolution studies, though  based
on only 4 absorbers. For absorbers with $13.1< \log \NHI< 14.0$, the
redshift density is $\sim 1.5\sigma$ lower than the mean of comparable low redshift
studies at the same resolution.   Absorbers with $\log \NHI \geq 12.5$ and 
a  subset using the same error-limited, $b<40$~\kms\ sample criteria as used
by \citet{Lehner07a}  both have a similar  redshift density to predictions
from numerical models by \citet{Paschos09} and  \citet{Dave99}.

3. The Doppler parameter distribution has a  mean, median and standard
deviation of 28, 27, 13~\kms , respectively.   There is no significant
evidence that line blending affects the distribution. If we choose a
restricted sample with errors $\leq 40$\% in $b$ and $\log \NHI$ and remove
absorbers with $b>40$~\kms  , the mean and standard deviation are 24, 24 and
8~\kms , respectively,  which is consistent both with the main \Lya\ system
sample from \citet{Lehner07a} and also with their similarly restricted data
set for  $b\leq 40$~\kms .

4. The \HI\ column density distribution to $\log \NHI=12.5$  is consistent
with a power law fit to a sample $7$ times larger covering $13.2<\log \NHI
<16.5$ from \citet{Lehner07a}.  If we correct our sample for the probability
of detection at $12.3<\log \NHI <12.5$, there is no evidence for a
significant break in the power law fit to $\log \NHI = 12.3$.

5. We find a total of 3 broad \Lya\ absorbers (BLAs) with $b>40$~\kms\ and
$\log \NHI/b \approx 10.9$, one of which is deemed reliable.  The redshift
densities are consistent with results from  hydrodynamical simulations by
\citet{Richter06a}.  The reliable BLA at $z=0.087632$ is $\approx
700$~\kms\  from an \OVI\  absorber in velocity space.  The $z=0.0902$ \OVI\
absorber is in a high galaxy density environment with a probability of being
in an equal or richer environment of  $P\approx 0.001$, based on galaxy
counts in the SDSS DR6.  The reliable BLA is in a marginally  overdense
environment as well ($P\approx 0.007$), and both may be embedded in the
same  structure.

6. We find evidence for weak clustering in the \Lya\ forest  at $3.4\sigma$
significance for  $\log N(\HI ) \geq 12.6$ from the two point correlation
function on a scale of $\Delta  v<1000$~\kms .   The limits obtained are
consistent with the clustering expected at that column density limit from
hydrodynamical simulations.

7. We find evidence for a void in the \Lya\ forest for $\log \NHI \simgt
12.6$ at $0.0901<z<0.1200$, with a random probability of occurrence
$P=0.006$.  The redshift interval does not cover  an unusually low galaxy
density environment. The relation between weak absorbers and galaxy
environment  associated with the 3C~273 and other low $z$ sight lines should
be addressed with deeper galaxy surveys than the SDSS.

8. A re-analysis of the $z=0.0053$ Virgo Cluster metal absorption system
using  both \FUSE\ and STIS data reveals velocity structure in the \HI\
absorption which is best fitted by  2 components.  The total \HI\ column
density is within $2\sigma$ of that used  for a previous analysis of the
physical properties of the system, whose results are  not significantly
changed by the adoption of a relatively weak second absorber.

3c~273 provides one of the best probes to very low column density limits at
low  redshift, and thus a unique probe of halo gas around galaxies in the
local universe. Further studies with COS and a deep, complete galaxy survey
will no doubt reveal further details of the galaxy-absorber relation.

{\bf Acknowledgments}
%\acknowledgements

This work was partially supported by the STIS IDT through the National
Optical Astronomical Observatories and by the Goddard Space Flight Center.
Based on observations obtained with the NASA/ESA Hubble Space Telescope,
which is operated by the Association of Universities for Research in
Astronomy, Inc., under NASA contract NAS 5-26555, and the NASA Far
Ultraviolet Spectroscopic Explorer.   We also used the NASA Extragalactic
Database (NED).  We thank  Dave Bowen, Charles Danforth, Rajib Ganguly,
Zoltan Haiman, Chris Howk, Jim  Lauroesch, Nicolas Lehner, Jon Loveday,  Joe
Meiring, Emma Ryan-Weber and the late Ervin Williger for useful discussions,
manuscript-reading and encouragement, and Eckart Janknecht, Tae-Sun Kim,
Jason Prochaska and Philipp Richter for discussions and assistance in
assembling comparison data from the literature. We appreciate helpful comments
from the referee for improving the presentation of this work.
GMW is acknowledges support
from U. Louisville startup funds, and is grateful for  hospitality from
Cambridge University,  E\"otv\"os University, Konkoly Observatory,
Observatoire de Paris-Meudon and Institut d'Astrophysique de Paris   during
the course of this work.

\appendix
\section{\bf A comparison with the line list from Danforth \& Shull (2008)}
\label{appendix-comparison}

\cite{Danforth08} published a study of a large sample of QSO absorbers,
including the line list for a data set analysed in detail in this  work. 
There are a number of differences in the reduction and analysis.
\citeauthor{Danforth08} used {\sc Calfuse 2.4} to reduce the \FUSE\ data,
whereas we used {\sc Calfuse 3.1.3} \citep{Dixon07}, which has many changes,
including a  significant  improvement in the wavelength calibration from
version 2.4. \citeauthor{Danforth08} normalised the data in 10~\AA\ segments
centred on IGM absorbers and binned the \FUSE\ and smoothed the STIS data by
3 pixels each,  whereas we used a mainly automated continuum-fitter with
manual adjustments around emission lines and did not bin or smooth.
\citeauthor{Danforth08} used a combination of apparent optical depth 
\citep{Sembach92}, curve of growth concordance plots and Voigt profile
fitting, whereas we mainly used Voigt profile fits and only used apparent
optical depth to establish lower limits for saturated metal transitions.
\citeauthor{Danforth08} tended to treat blends which they could not separate
unambiguously as single absorbers,  whereas our normal procedure is to add
components when the existing  Voigt profile fit has a probability of
matching the data of $P<0.01$.

\begin{figure}
%\plotone{lya2pc.n125170.dv500.vsm27.06jan.mc1e4.ps}
%\epsscale{1.1}
\includegraphics[width=185mm]{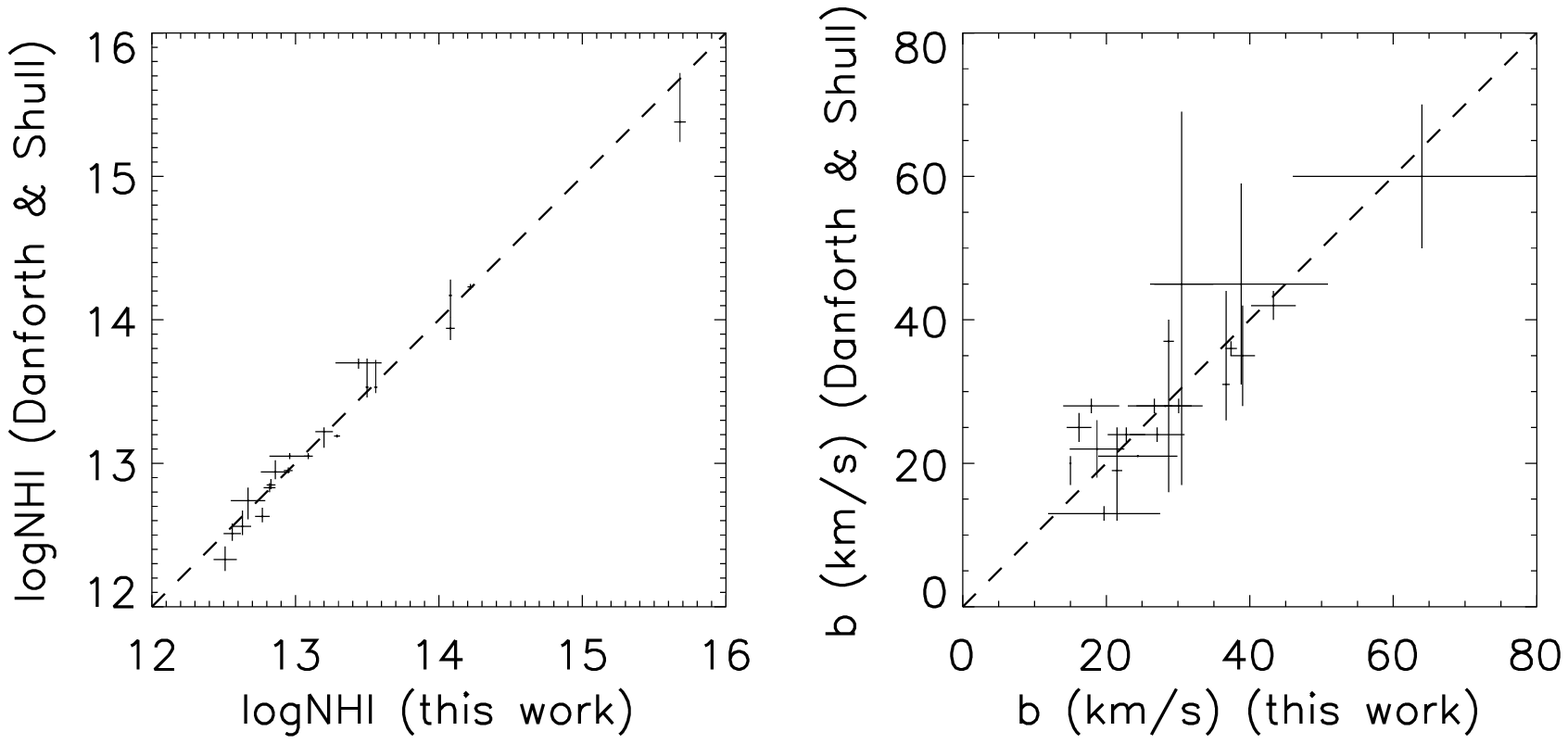}
\caption{Comparison of \HI\ column density and Doppler parameter for 20 
absorbers in common between \citet{Danforth08} and this work, with $1\sigma$
error bars.   Column densities have been summed in the case of fitted
complexes in this work spanning $\Delta v\simlt 70$~\kms . However, column
density errors for this study have been retained for the highest \NHI\
component, as are the Doppler parameters and errors.  Fits are generally 
consistent within $1\sigma$ errors.  Differences are qualitatively
understandable in that \citeauthor{Danforth08} have slightly higher column
densities and Doppler parameters due to their not using as  fine component
structure as in this work.
\label{fig:comparedanforth}
}
\end{figure}

We matched absorbers from the \citeauthor{Danforth08} list with ours, and
combined apparently blended absorbers in our line list  in velocity space
(up to $\Delta v\sim 70$~\kms ) by summing their \hi\ column densities.  We 
display the comparison for 20 systems in common in
Fig.~\ref{fig:comparedanforth}. For the plot we use the errors in $\log
\NHI$ for our fits, and also the Doppler  value and corresponding errors,
for the highest $\NHI$ component. The \hi\ column densities show a good
match within $1\sigma$ errors.  The  Doppler parameters show more scatter,
which likely arises from the finer structure  decomposition in this study,
but are still largely consistent within $1\sigma$ errors.

Absorbers in one list but not the other, or potentially suspect absorbers,
are as follows.

$z=0.002630:$  This is a broad line which \citeauthor{Danforth08} do not
list. We need this component for a proper fit to the Galactic \Lya\ profile,
which we fitted simultaneously with the IGM components in the  wing. It is
below the minimum redshift for any of our statistical analyses.

$z=0.007588:$ This feature is included by \citeauthor{Danforth08} as well. 
It  is in the wing of Galactic \HI , and shows a double minimum. The complex
has $\sim 3\sigma$ significance overall.  We include it due to its
contribution to a fit to the Galactic \HI\ complex,  but note that the
profile is atypical and may indicate a blend of two weaker features.

$z=0.054399:$ This is a weak line which \citeauthor{Danforth08} do not list.
It is below our 80\% probability of detection threshold of $\log \NHI = 12.5$, 
but is
used for the \hi\ column density extrapolation to $\log \NHI = 12.3$.

$z=0.06051:$ This is a weak line from \citeauthor{Danforth08}
which we do not list due to its low significance.  It contains a small noise 
spike halfway
through the feature, and 
has a square profile and flat minimum.

$z=0.064078:$ This is a shallow, moderately broad feature which 
\citeauthor{Danforth08} do not list. We find satisfactory constraints on the
column density and Doppler parameter  such that we use it in our statistical
samples.

$z=0.073738$ and $0.074286:$ These are broad, shallow feature which 
\citeauthor{Danforth08} do not  list. Their high Doppler parameters of
$b=64\pm 15$~\kms\ and $b=51\pm 14$~\kms\ make  us omit them from our
restricted sample
analysis, to be consistent with the \cite{Lehner07a} $b<40$~\kms\ sample 
criteria. The $z=0.073738$ system is unlikely to be Galactic 
\OIstar , since its redshifted corrected position is just outside the
\OI ~1032 profile. There may be a better case for  O\,{\sc I}$^{**}$ being
the $z=0.07429$ feature, since it falls in the ground state profile, and 
there is possibly \OIstar\ in the red wing of Si\,{\sc III}~1304.
However, neither  agree particularly well in position with C\,{\sc II}$*$,
so unless they would be at  significantly different velocities (which is
not suggested by the relative positions of the \CII / \OI\ centroids) then
we doubt that the  \OIstar ,  O\,{\sc I}$^{**}$ identifications apply.

$z=085882:$ This is a broad,  shallow feature which \citeauthor{Danforth08}
do not list. It is below our 80\% probability of detection threshold of
$\log \NHI = 12.5$,  but is used for the \hi\ column density extrapolation
to $\log \NHI = 12.3$.

$z=0.109141:$ This is a  shallow, round-bottomed feature which 
\citeauthor{Danforth08} do not list. We use it in our extrapolation of the
\hi\ column density distribution to $\log  \NHI=12.3$, but in no other
statistical sample.

$z=0.13943$  Danforth et al. list it, but we find it less plausible. It is 
assymetrical and contains a noise spike.  The significance is $\sim
3\sigma$, and depends  strongly  on the integration bounds.

$z=0.141676$ and $z=0.150204:$  These are  broad, shallow features  which
\citeauthor{Danforth08} do not list.  They have $\sim 3\sigma$ and 
$4.5\sigma$ significance, respectively. They fall on \Lya\ emission and
within our definition of  the proximity effect zone ($\Delta v<5000$~\kms ),
and are also below our 80\%  probability detection limit.   We do not use
them for any analysis.

$z=0.157787:$  This is a distinct absorber with a hint of an extended red
wing,  noted in \citet{Tripp08} in association with \OVI .  We  fitted the
feature with one component.  \citeauthor{Danforth08} do not list it, 
presumably because it falls within their definition of  the proximity effect
zone ($\Delta v<1500$~\kms ).

\clearpage

\clearpage

\end{document}